\documentclass[fleqn,usenatbib]{mnras}

\usepackage[T1]{fontenc}

\DeclareRobustCommand{\VAN}[3]{#2}
\let\VANthebibliography\thebibliography
\def\thebibliography{\DeclareRobustCommand{\VAN}[3]{##3}\VANthebibliography}

\usepackage{graphicx}	
\usepackage{amsmath}	
\usepackage{amssymb}	

\usepackage{tabularx}

\newcommand{\bs}{\boldsymbol}
\newcommand{\percent}{\,\mathrm{per\,cent}}

\usepackage[table]{xcolor}
\definecolor{lightgray}{gray}{0.9}
\definecolor{darkred}{rgb}{0.76, 0.23, 0.13}
\definecolor{pink}{rgb}{0.96, 0.76, 0.76}

\usepackage{newtxtext,newtxmath}

\title[Classification of LPVs with BP/RP spectra]{Hunting for C-rich long-period variable stars in the Milky Way's bar-bulge using unsupervised classification of Gaia BP/RP spectra}

\author[J.~L. Sanders \& N. Matsunaga]{Jason L. Sanders$^{1}$\thanks{jason.sanders@ucl.ac.uk (JLS)}
and
Noriyuki Matsunaga$^{2}$
\\
$^{1}$Department of Physics and Astronomy, University College London, London WC1E 6BT, UK\\
$^{2}$Department of Astronomy, School of Science, The University of Tokyo, 7-3-1, Hongo, Bunkyo-ku, Tokyo 113-0033, Japan
}

\date{Accepted XXX. Received YYY; in original form ZZZ}
\pubyear{2022}

\begin{document}
\label{firstpage}
\pagerange{\pageref{firstpage}--\pageref{lastpage}}
\maketitle

\begin{abstract}
The separation of oxygen- and carbon-rich AGB sources is crucial for their accurate use as local and cosmological distance and age/metallicity indicators. We investigate the use of unsupervised learning algorithms for classifying the chemistry of long-period variables from Gaia DR3's BP/RP spectra. Even in the presence of significant interstellar dust, the spectra separate into two groups attributable to O-rich and C-rich sources. Given these classifications, we utilise a supervised approach to separate O-rich and C-rich sources without BP/RP spectra but instead given broadband optical and infrared photometry finding a purity of our C-rich classifications of around $95$ per cent. We test and validate the classifications against other advocated colour--colour separations based on photometry. Furthermore, we demonstrate the potential of BP/RP spectra for finding S-type stars or those possibly symbiotic sources with strong emission lines. Although our classification suggests the Galactic bar-bulge is host to very few C-rich long-period variable stars, we do find a small fraction of C-rich stars with periods $>250\,\mathrm{day}$ that are spatially and kinematically consistent with bar-bulge membership. We argue the combination of the observed number, the spatial alignment, the kinematics and the period distribution disfavour young metal-poor star formation scenarios either in situ or in an accreted host, and instead, these stars are highly likely to be the result of binary evolution and the evolved versions of blue straggler stars already observed in the bar-bulge.
\end{abstract}



\section{Introduction}
Stars of masses between $\sim0.8$ and $\sim8M_\odot$ will eventually pass through the asymptotic giant branch (AGB) phase, the final nuclear burning stage, characterised by high luminosity and large cool convective envelopes \citep{Herwig2005}. AGB stars are significant components of a galaxy's stellar population both in terms of luminosity but also in terms of their contributions to the nucleosynthetic build-up of elements in their galaxy \citep{Karakas2014,Kobayashi2020}. They are also typically long-period variables (LPV) with the Mira variables representing the highest amplitude AGB pulsators typically attributed to fundamental mode pulsations. During these pulsation phases, AGB stars lose significant quantities of their mass producing large amounts of dust \citep{Hofner2018}. The chemistry of the AGB envelope and the properties of the circumstellar dust are determined by the carbon to oxygen ratio, C/O, which in turn affects the subsequent evolutionary properties of the star \citep[see][for thorough reviews of carbon stars]{Wallerstein1998,LloydEvans2010}. The sub-dominant element will be bound up almost entirely in CO leaving the dominant element to form carbon- or oxygen-rich molecules such as C$_2$ and CN, and TiO and SiO respectively. When C/O $\approx1$ the AGB star will be an S-type star exhibiting intermediate chemistry with a mixture of carbon- and oxygen-rich species and characteristically ZrO.

The C/O ratio of an AGB star is altered by the third dredge-up bringing newly-synthesised carbon to the outer atmosphere. The strength of the dredge-up and its subsequent impact on the C/O ratio is a function of both the metallicity and age/mass of the AGB star \citep{Karakas2014b,Hofner2018}. Loosely, higher mass stars have stronger dredge-up episodes, and lower metallicity stars have lower levels of oxygen in their atmosphere so are more readily diluted by dredged-up carbon. In this way, the fraction of C-rich to O-rich AGB stars can be a useful indicator of the age and/or metallicity of a stellar population \citep{Brewer1996,Boyer2013}. For example, in the Milky Way disc, the fraction of C-rich to O-rich AGB stars increases with radius \citep{Blanco1984,Ishihara2011} which could be a reflection of the metallicity gradient and/or the inside-out growth of the disc. Due to its predominantly old, metal-rich population \citep{Bensby2013,Catchpole2016,Bovy2019,Hasselquist2021}, the Milky Way's bar-bulge is expected, and largely observed, to host relatively few C-rich LPV stars. \cite{Matsunaga2017} discovered five C-rich Mira variable stars with magnitudes consistent with bar-bulge membership and argued that the previously-discovered symbiotic C-rich Mira variable from \cite{Miszalski2013} was more consistent with foreground disc membership because the 2MASS data was taken around the light curve minimum. Although \cite{Matsunaga2017} favour binary evolution as the production mechanism for these C-rich variables, it is also possible they are the result of single star evolution and arise from either accretion or recent in-situ metal-poor star formation. The occurrence of dwarf carbon stars \citep[e.g.][]{Whitehouse2018}, CH stars and carbon-enhanced metal-poor stars rich in s-process elements \citep{Koch2016,ArentsenPIGS} demonstrates that binary evolution can significantly alter the C/O ratio of some stars, particularly those that are metal-poor \citep{DeMarco2017}. Indeed \cite{Azzopardi1988,Azzopardi1991} found a population of C-rich giants towards the Galactic bulge that are too faint to be AGB stars in the bulge but could instead be lower luminosity red giant branch stars in the bulge that are the result of binary evolution \citep[or plausibly associated with the Sagittarius dwarf spheroidal galaxy,][]{Ng1997,Ng1998}. Hunting for C-rich stars in the Galactic bulge is a useful archaeological probe for discovering accreted or young in-situ populations, but disentangling the different formation channels to reveal details of the evolutionary history of the Galaxy requires the complement of spectroscopic and kinematic analysis now possible with Gaia.

Observations in the Large Magellanic Cloud \citep[LMC, e.g.][]{GlassLloydEvans1981,Iwanek2021b} have demonstrated that LPVs reside along a series of period--luminosity relations believed to be associated with different pulsation modes \citep{Wood2000}. In particular, the high-amplitude Mira variables lie along a tight sequence in the dust-insensitive Wesenheit indices vs. period \citep{Soszynski2009}. For this reason, Mira variables are increasingly seeing use as distance tracers for structure within the Milky Way and Local Group \citep{Catchpole2016,Deason2017,Grady2020,Semczuk2022}, and also as an alternative distance ladder calibrator for measurements of $H_0$ \citep{Huang2018,Huang2020}. However, in general, O-rich and C-rich Mira variables are known to lie along separate period--luminosity sequences with the O-rich Mira variables typically obeying tighter period--luminosity relations \citep{GlassLloydEvans1981,Feast1989,Ita2004,Groenewegen2004,Fraser2008,Riebel2010,Ita2011,Yuan2017,Yuan2018,Bhardwaj2019,Iwanek2021b}. This appears to be in large part due to the differing amounts of circumstellar dust \citep{Ita2011}. Careful characterisation and separation of the two types are essential for precision work both in the cosmological and Local Group setting.

\begin{figure*}
    \centering
\includegraphics[width=\textwidth]{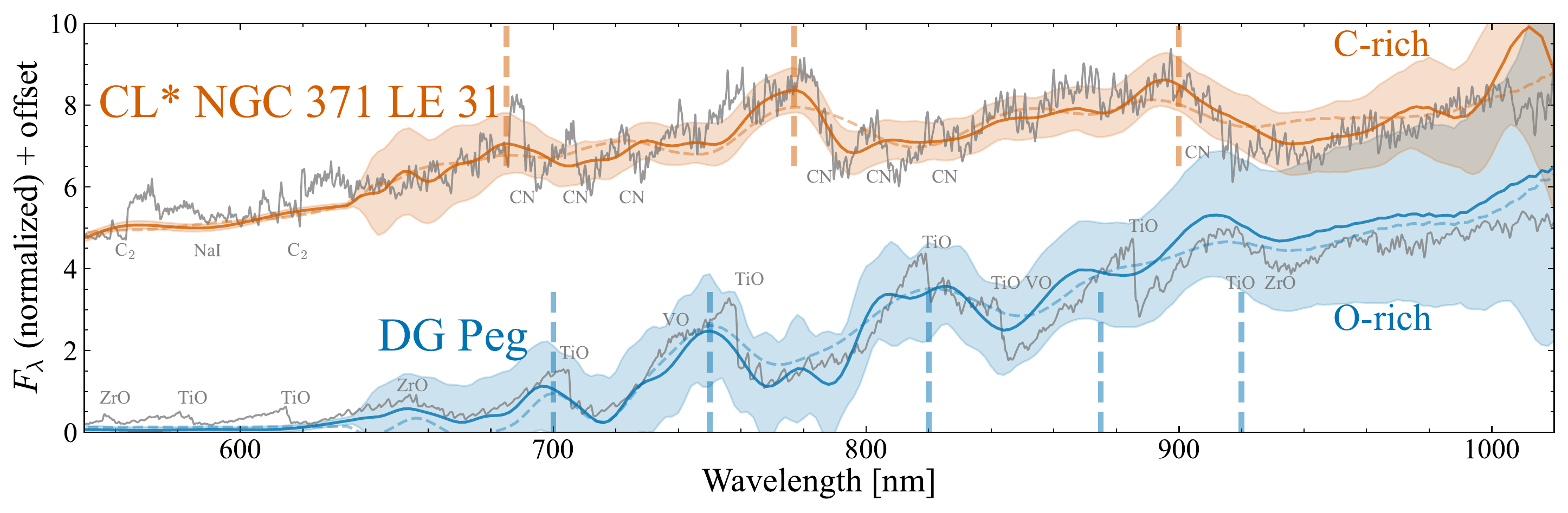}
    \caption{Two example long-period variable star BP/RP spectra (with the $\pm1\sigma$ bracket) compared to spectra from the X-Shooter library \protect\citep[XSL, smoothed to $0.36\,\mathrm{nm}$,][]{Chen2014,Gonneau2021,Verro2022}. The dashed line shows the truncated BP/RP spectra. All spectra have been normalized by the integral of the XSL spectra from $350$ to $1020\,\mathrm{nm}$ and the C-rich spectrum has been offset vertically for clarity. DG Peg is an oxygen-rich Mira variable star (spectral type M4e) whilst CL* NGC 371 LE 31 is a carbon-rich semi-regular variable in the SMC cluster NGC 371. We have marked the peaks arising from the molecular band heads (as labelled), which are easily distinguishable in the BP/RP spectra (primarily TiO for O-rich, CN for C-rich).}
    \label{fig:example_spectra}
\end{figure*}

Traditionally, carbon stars have been identified by their C$_2$ Swan bands and CN bands in objective-prism plates \citep{Secchi1868,Nassau1964,MacConnell1988,Aaronson1990}. In more recent years, the separation of  O-rich and C-rich AGB stars has been performed on large samples using infrared photometry taking advantage of the silicate bands in the O-rich spectra at $\sim9$ and $\sim18\,\mu\mathrm{m}$ compared to the SiC band at $\sim11\,\mu\mathrm{m}$ e.g. AKARI \citep{Ishihara2011}, WISE \citep{Lian2014,Nikutta2014,SuhHong2017}, Spitzer \citep{Kastner2008,Groenewegen2018} and MSX \citep{Lewis2020a,Lewis2020b} guided by the more detailed infrared spectroscopic view from the IRAS LRS, the ISO SWS and the Spitzer IRS instruments \citep{Olnon1986,Kraemer2002}. \cite{Soszynski2009} separated O-rich and C-rich sources in the LMC using the Wesenheit $W_{I,V-I}$ vs. $W_{Ks,J-Ks}$ diagram where the two Wesenheit indices are $W_{I,V-I}=I-1.55(V-I)$ and $W_{Ks,J-Ks}=K_s-0.686(J-K_s)$. In a similar vein, \cite{Lebzelter2018} proposed a scheme for separating O-rich and C-rich sources in the LMC using a combination of Gaia and 2MASS broadband photometry. This is highly desirable due to the all-sky availability of this photometry. These authors advocated for a separation in the `colour--magnitude' diagram of $W_\mathrm{RP,BP-RP}-W_{Ks,J-Ks}$ vs. $K_s$ where $W_\mathrm{RP,BP-RP}=G_\mathrm{RP}-1.3(G_\mathrm{BP}-G_\mathrm{RP})$. There is relatively weak curvature of the O-rich/C-rich separation line in this space meaning an analogue of this separation using only $W_\mathrm{RP,BP-RP}-W_{Ks,J-Ks}$ can also in theory be used for non-LMC stars. These Wesenheit diagrams are effective as, in the optical, oxygen-rich AGB stars have a set of TiO bands \citep[also ZrO and VO, see figure 4 of][for identification of the lines, see also \citealt{VanEck2017} and \citealt{Yao2017}]{Lancon2000} whilst the carbon-rich stars have a set of Swan C$_2$ bands (at $<600\,\mathrm{nm}$) and CN bands (at $>700\,\mathrm{nm}$) \citep[e.g. figure 7 of][]{Lancon2002}. Representative O-rich and C-rich LPV spectra from the X-Shooter Library \citep{Chen2014,Gonneau2021,Verro2022} are shown in Fig.~\ref{fig:example_spectra} where these bands are clearly visible. Both the locations and separations between the bands are distinct for the two types of star leading \cite{Lebzelter2022} to develop a classification on the basis of peak separation in pseudo-wavelength (\texttt{median\_delta\_wl\_rp}) in the Gaia DR3 RP spectra. However, \cite{Lebzelter2022} report that the classification scheme performs poorly for low signal-to-noise spectra and/or highly-extincted sources and suggest only trusting the C-rich classification if $7 < \texttt{median\_delta\_wl\_rp} < 11$ and $G_\mathrm{BP} < 19$.

Here we investigate the performance of an unsupervised O-rich/C-rich classification scheme using the Gaia third data release (DR3) BP/RP spectra \citep[also called XP spectra,][]{Carrasco2021,DeAngeli2022,Montegriffo2022}. \cite{Lucey2022} has already demonstrated the power of utilising a supervised classification on the BP/RP spectra for the identification of carbon-enhanced metal-poor stars. Unsupervised classification seeks to find similarities between presented training examples such that clusters of `similar' data can be assigned labels. Numerous clustering algorithms exist but often the challenge is representing the dataset in a space that is amenable to clustering. In many cases, the data are of high dimensionality making clustering algorithms very computationally expensive, or the natural clustering is along complex surfaces in the high-dimensional space. One solution to both of these problems is to project the data to a lower dimensional space that encodes as much information from the higher dimensional space as possible. For example, principal component analysis (PCA) seeks to find the most informative linear combinations of the higher dimensional space. However, it is inappropriate for significantly reducing the dimensionality of the data as it is limited to only considering linear combinations of the input data dimensions whilst often clusters lie along highly non-linear surfaces.

Several algorithms have sought to circumvent this limitation. For example, t-SNE \citep[t-stochastic neighbour embedding,][]{tSNE} first finds the similarities between the data points in the high dimensional space $\bs{x}$ using a Gaussian kernel and then attempts to find the low dimensional projection $\bs{y}$ that minimizes the Kullback--Leibler (KL) divergence between the higher dimensional similarity distribution and that of the lower dimensional representation assuming the lower dimensional similarity distribution follows a Student t distribution. t-SNE has found considerable use for astrophysics applications, in particular in the analysis of spectroscopic datasets \citep[e.g.][]{Traven2017,Anders2018}. One disadvantage of the t-SNE algorithm is that it only preserves a sense of distance (metric) between local points in the lower dimension space. Additionally, the original implementation was computationally expensive for large datasets (mostly due to having to construct the $N$ by $N$ high-dimensional similarity distribution) and in practical applications is often combined with an initial PCA to an intermediate dimensionality dataset. UMAP \citep[Uniform Manifold Approximation and Projection,][]{McInnes2018} was designed to solve both of these issues with t-SNE. Instead of utilizing the KL divergence, UMAP uses the cross-entropy which ensures $|\bs{y}_i-\bs{y}_j|\rightarrow\infty$ as $|\bs{x}_i-\bs{x}_j|\rightarrow\infty$ (whilst for the KL divergence $|\bs{y}_i-\bs{y}_j|$ can take any value as $|\bs{x}_i-\bs{x}_j|\rightarrow\infty$). Other computational/algorithmic improvements have enabled significant speed-ups for UMAP compared to t-SNE although in essence, the algorithms share significant similarities and the t-SNE implementation from \cite{openTSNE} is competitive with UMAP in terms of computational time. Despite its relatively recent creation, the UMAP algorithm has already found use in astrophysics applications \citep{Reis2019,Kim2022,Grondin2022}.

Here we investigate the use of these unsupervised classification algorithms (UMAP and t-SNE) for determining the chemistry of AGB stars using their Gaia DR3 BP/RP spectra.
We begin by describing the BP/RP spectra and our chosen unsupervised classification scheme in Section~\ref{sec:unsupervisedClassification}. BP/RP spectra are unavailable for stars with $G>17.65$ although there are many identified LPVs from Gaia fainter than this limit. We, therefore, extend our unsupervised classification to the fainter objects with a supervised scheme using Gaia and 2MASS photometry in Section~\ref{sec:supervisedClassification}. We validate our results by inspecting previously employed colour-colour diagrams for separating C-rich and O-rich sources in Section~\ref{sec::validation}. Finally, we use the new classification results to search for C-rich variables in the Galactic bulge in Section~\ref{sec::bulge}.

\section{Unsupervised classification of O-rich/C-rich long-period variables}\label{sec:unsupervisedClassification}

Our primary data source is the Gaia DR3 long-period variable candidate catalogue \citep{Lebzelter2022} as part of the full third Gaia data release \citep[][Vallenari et al. in prep.]{Gaia1} that complements the astrometric results from the early third Gaia data release \citep{GaiaEDR3}. This is the second version of the long-period candidate table from Gaia after the first catalogue of \cite{Mowlavi2018} based on Gaia DR2 photometry. Gaia's long-period variable processing consists of two pipelines: a generic classification pipeline for assigning classes to all variables \citep{Rimoldini2019} and then the specific object study (SOS) for the stars classified as long-period variables by the first stage (in addition to a very low number of additional likely LPVs largely classified as symbiotic stars in the classification pipeline). Additional quality cuts are also performed on the stars processed by the SOS on the basis of colour, high $G$-band variability, number of epochs and valid astrometry. In total there are $1,720,588$ stars in the Gaia DR3 long-period variable candidate catalogue of which $392,240$ have reported periods.

The BP/RP spectra from Gaia DR3 \citep{Carrasco2021,DeAngeli2022,Montegriffo2022} are low resolution ($R=\lambda/\delta\lambda\approx25-100$) and together the blue and red photometers (BP/RP) cover the optical range from $330$ to $1050\,\mathrm{nm}$. The spectra are provided for $\sim220$ million stars in Gaia with $G<17.65$. Despite their low resolution, several studies have demonstrated that the information content is sufficient to measure bulk spectroscopic parameters \citep[$T_\mathrm{eff}$, $\log g$, {$[\mathrm{M}/\mathrm{H}]$},][]{Liu2012,Witten2022,Xylakis2022,Andrae2022,Fouesneau2022,Creevey2022APSIS,Belokurov2022,Rix2022} although detailed chemical abundances are likely too challenging \citep{Gavel2021}. However, particularly in cooler stars, the presence of molecular features in these spectra is detectable \citep{Lucey2022} enabling more accurate metallicity determinations and detailed carbon abundances. For each star, the BP/RP spectra are provided as two sets of coefficients (with associated covariance matrices) from which both the internally- or externally-calibrated spectra can be reconstructed on a wavelength grid of choice through multiplication with a set of basis functions (times a mixing term for connecting the BP and RP spectra at their interface). This can be done using the \textsc{GaiaXPy} python package\footnote{Available from \url{https://gaia-dpci.github.io/GaiaXPy-website/}. We use v1.1.2 \url{10.5281/zenodo.6642313}.} \citep{Montegriffo2022}. The basis functions are linear combinations of Hermite polynomials chosen through a PCA/SVD on a set of BP/RP calibrators. In this way, the first coefficient contains the most information for the average calibrator star and higher-order coefficients provide ever-decreasing corrections to this average calibrator star's spectrum. It is expected that stars that are significantly different to the average calibrator star in the full dataset will not necessarily follow this hierarchy. The Gaia DR3 data also provides a truncation order (\texttt{xp\_n\_relevant\_bases}) for each star indicating the last coefficient that is significant compared to the noise. Of the $1,720,588$ Gaia DR3 LPVs, $1,205,121$ have BP/RP spectra.

In Fig.~\ref{fig:example_spectra} we display two spectra from the X-Shooter Library \citep{Chen2014,Gonneau2021,Verro2022}: DG Peg is an O-rich Mira variable star (Gaia DR3 $1768290812321736576$ with a $147$ day period and $G=11.1\,\mathrm{mag}$) whilst CL* NGC 371 LE 31 is a semi-regular variable star \citep[although with a high $I$ amplitude of $\approx0.69\,\mathrm{mag}$]{Soszynski2009} in the Small Magellanic Cloud open cluster NGC 371 (Gaia DR3 $4690506059969811968$ with a $301$ day period and $G=16\,\mathrm{mag}$). We have normalized the spectra by the integral $\int\mathrm{d}\lambda\,\lambda F_\lambda$ over $350$ to $1020$ nm and the C-rich spectrum is offset vertically for clarity. We also display the Gaia DR3 BP/RP spectra constructed from the coefficients on a $2\,\mathrm{nm}$ grid (using \textsc{GaiaXPy}) along with the $\pm1\sigma$ band implied by the covariance matrix for the coefficients. The BP/RP spectra have been divided by the same normalization factor as the X-Shooter spectra. There is a very good correspondence between the BP/RP spectra and the X-Shooter spectra showing that both are well absolutely calibrated. We further display the BP/RP spectra constructed by truncating the spectra as dashed lines (the number of relevant RP bases are $2$ and $4$ for DG Peg and CL* NGC 371 LE 31 respectively) We see in these cases the truncated spectra largely do a good job of capturing the features we are interested in here. From Fig.~\ref{fig:example_spectra} it is clear that the O-rich and C-rich spectra display very different sets of features that are well captured in the BP/RP spectra. The O-rich DG Peg spectrum shows a series of TiO bands at 
$705.4$, $758.9$, $819.4$ $843.2$, $885.9$, and $920.9\,\mathrm{nm}$ \citep{Bobrovnikoff1933,Sharpless1956,Murset1999}
that give rise to characteristic peaks at $700$, $750$, $820$ and $920$ nm with the band head at $885.9$ nm giving rise to a peak around $875$ nm that often blends with the $920$ nm peak in the BP/RP spectra. There are a series of other band heads bluer than $700$ nm also due to TiO but in general, these are hard to identify in the BP/RP spectra due to the lower flux. Figure 4 of \cite{Lancon2000} labels the TiO transitions responsible for these characteristic peaks/troughs in a cool and a warm Mira spectrum. In the cooler spectra, VO features appear at $\sim740$, $\sim790$ and $\sim860$ nm broadening the neighbouring TiO features (there is the suggestion of this in the double minimum structure at $780\,\mathrm{nm}$ where the right minimum is due to VO absorption). The C-rich spectrum however has a set of three peaks at $683$, $778$ and $900$ nm due to a series of CN features \citep[see e.g.][]{Gonneau2016}. For distinguishing O-rich and C-rich stars, the region between $750$ and $850$ nm is particularly clear where the CN band head sits at the minimum between two TiO band heads. It is quite clear from this example that BP/RP spectra have the capability to distinguish between O-rich and C-rich stars \citep[as already evidenced by][]{Lebzelter2022}.

As we are dealing with solely variable stars, it is worth considering how variability alters our interpretation of the Gaia data. The BP/RP spectra are the average of Gaia's observations of each source over many transits (typically $20-40$). For variable stars, we therefore approximately observe the properties of the stars averaged over period (Gaia DR3's observing window is $34$ months so most LPVs have at least one period of observations). During the pulsation of long-period variables, the temperature of their envelopes varies giving rise to different balances of molecular species (as well as varying emission line ratios e.g. \citealt{Yao2017} although these are less important for our work). In O-rich Mira variables, TiO is only present at low levels at maximum brightness (temperature) but its production strongly increases towards lower temperatures giving rise to the high amplitudes in the visual bands \citep{Reid2002}. In C-rich Mira variables, C$_2$ and CN are formed at higher temperatures deeper in the atmosphere and so their relative abundance does not change significantly during the pulsation cycle \citep{Lancon2000}. Figure 3 of \cite{Lebzelter2022} shows the RP spectra of the O-rich star T Aqr and the C-rich star RU Vir at the individual observing epochs where it is clear that for the O-rich star the TiO bands reduce in depth at peak brightness (particularly the band at $\sim900\,\mathrm{nm}$ or $\sim40$ in pseudo-wavelength units) whilst the depth of the C-rich bands remains quite similar across all epochs. Aliasing may arise as an issue for stars with periods around $190$ or $380$ day (awkward periods for Gaia's scanning law) where the BP/RP spectra may only be averaging over similar phases.

\begin{figure*}
    \centering
\includegraphics[width=\textwidth]{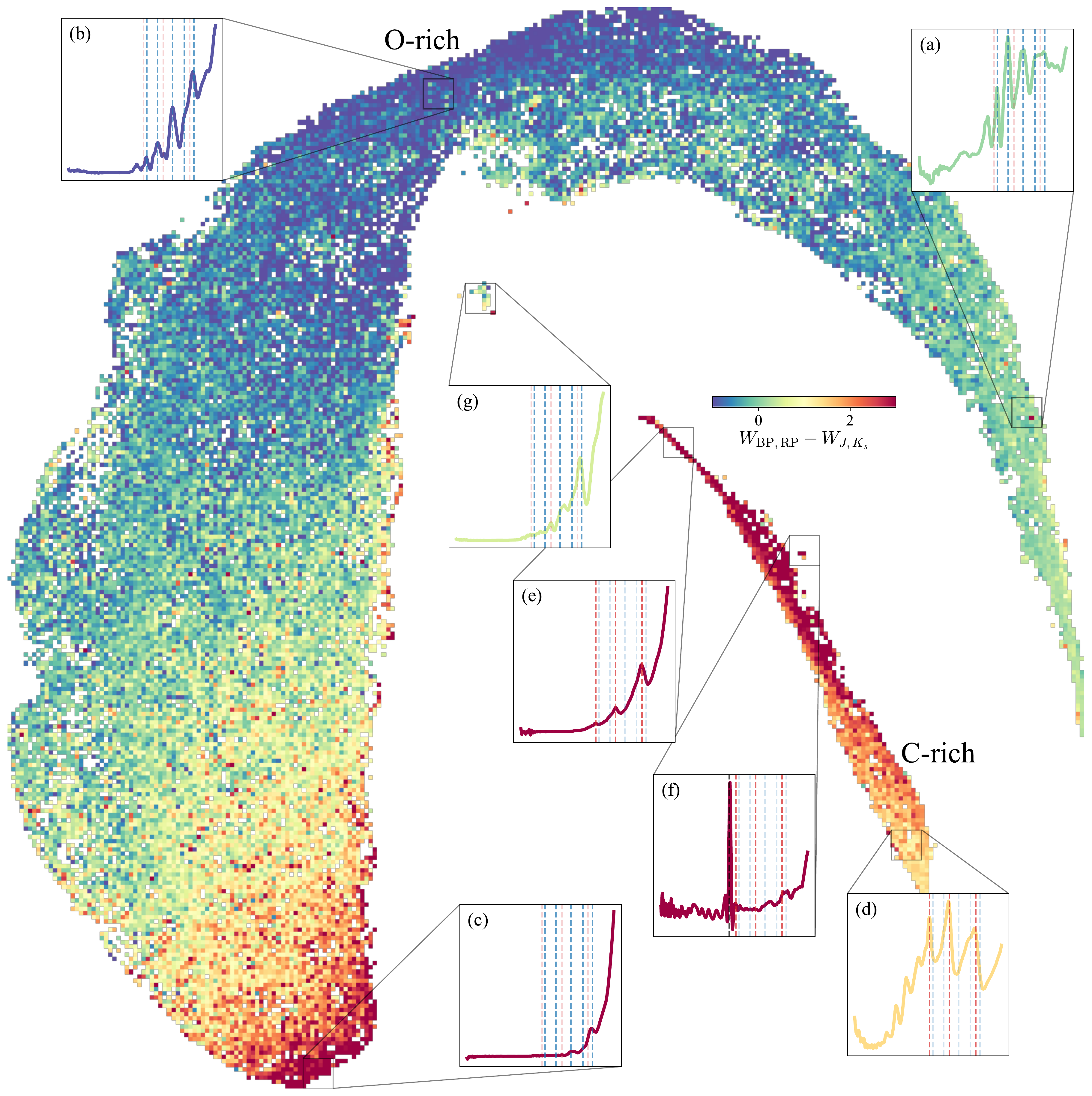}
    \caption{Two-dimensional UMAP projection of the BP/RP spectra coefficients for Gaia DR3 high-amplitude long-period variables. Each pixel is coloured by the mean Wesenheit colour $W_\mathrm{RP,BP-RP}-W_{Ks,J-Ks}$ advocated as a good metric for separating O-rich and C-rich stars by \protect\cite{Lebzelter2018}. The upper \emph{crescent} is populated by O-rich stars whilst the lower \emph{spur} is populated by C-rich stars. The insets show the median $\lambda^{-2}F_\lambda$ BP/RP spectrum between $336$ and $1020$ nm in each marked UMAP bin with the features from Fig.~\ref{fig:example_spectra} marked as dashed vertical lines (blue for O-rich, red for C-rich). As we move along the crescent from right to left (from (a) to (c)), the O-rich stars broadly become redder due to a combination of effective temperature and extinction variation. Moving vertically along the C-rich sequence from (d) to (e), the stars get redder. The small island (f) to the right of the C-rich sequence has significant H$\alpha$ emission (marked as black dashed) whilst the small island (g) off the centre of the crescent has a mixture of O-rich and C-rich features attributable to S-type stars.}
    \label{fig:umap}
\end{figure*}

\subsection{High-amplitude variable selection}
Our primary interest is the high-amplitude Mira variable stars in Gaia's LPV candidate catalogue. For this reason, we limit our analysis in the main body of this work to a high-amplitude subset of the Gaia DR3 LPVs. In Appendix~\ref{appendix::full} we show the results of running the approach on all Gaia LPVs. It should be noted that sources in the Gaia DR3 LPV catalogue are not assigned periods and amplitudes if the photometry is likely contaminated. In turn, this possibly means the BP/RP spectra are contaminated. Many of these sources are faint and in the crowded midplane and it is more of an issue for lower amplitude variables. In our analysis, we have opted to process all the stars without consideration of any data quality cuts under the proviso that any subsequent analysis will consider data quality more explicitly. For example, typical fields used for selecting high-quality data are the BP/RP photometric excess \citep[although as discussed by][the BP/RP photometric excess is a less useful quality cut for red variable sources]{Riello2021} and the fraction of contaminated transits. A further consideration is that some fraction of the classified LPVs may be contaminating young stellar objects \citep[YSOs, see the discussion in][for the Gaia DR2 LPV catalogue which likely still applies for the DR3 data]{Mowlavi2018}. However, some basic parallax cuts can approximately remove this contaminant.

We select all LPVs with peak-to-peak semi-amplitudes\footnote{We use $\Delta$ to denote the semi-amplitude throughout this work.}, $\texttt{amplitude}$, $>0.32\,\mathrm{mag}$ from the Gaia DR3 LPV candidate catalogue \citep{Lebzelter2022}. In Appendix~\ref{section::amplitude_comparison} we compare the different amplitude measures for these stars from Gaia DR3. This amplitude cut includes Mira variable stars which are typically defined as having $\Delta V>1.25$ and for which \cite{Grady2019} advocate a cut of $\Delta G>0.433\,\mathrm{mag}$. At the lower amplitude end, there will also be some semi-regular variables. Furthermore, around $190$ and $380$ day periods \texttt{amplitude} can significantly overestimate the true amplitude due to poor phase coverage related to Gaia's scanning law. However, for the majority of stars, $\texttt{amplitude}$ agrees with amplitude estimates from the scatter of the photometric data points. There are $99,212$ stars satisfying this amplitude cut of which $79,944$ have BP/RP spectra.

\subsection{Default procedure}
We use the BP/RP coefficients normalized to the first RP coefficient as our input data (note we are here ignoring uncertainties in the coefficients due to the limitations of the UMAP and t-SNE approaches). Our default setup is to run UMAP on the entire ($2\times55-1=109$) set of normalized coefficients to reduce it down to a two-dimensional projection. For UMAP there are two key hyperparameters, \texttt{min\_dist} and \texttt{n\_neighbors} setting the minimum distance between points in the low-dimensional space and the size of the neighbourhood used for finding the local manifold respectively. We have found \texttt{min\_dist} $=0.05$ and \texttt{n\_neighbors} $=15$ produces results that cleanly separate the dataset into two distinct groupings. After projection, we z-score the output and then run DBSCAN \citep{DBSCAN} to cluster the data with a standard maximum distance $\epsilon=0.09$ (although simpler linear cuts would also suffice).

The results of our procedure are shown in Fig.~\ref{fig:umap}\footnote{Note whenever we display a UMAP (or t-SNE) projection we have opted to drop the tick labels as the absolute values are unimportant.}. We see that in the two-dimensional UMAP space the high-amplitude variables split into two clear populations: a \emph{crescent} and a \emph{spur}. Fig.~\ref{fig:umap} is coloured by the mean Wesenheit colour $W_\mathrm{RP,BP-RP}-W_{Ks,J-Ks}$ which was advocated by \cite{Lebzelter2018} as a good space to separate O-rich and C-rich stars in the LMC. We see indeed that the spur has mostly $W_\mathrm{RP,BP-RP}-W_{Ks,J-Ks}>1$ so we identify it with C-rich sources whilst the crescent has mostly $W_\mathrm{RP,BP-RP}-W_{Ks,J-Ks}<1$ so we identify it with O-rich sources. Note however that at the far left end of the crescent the stars are redder and as red as the C-rich spur. 

To further validate our assignment of O- and C-rich to the crescent and spur, we have plotted a series of median BP/RP spectra in the insets (we divide by $\lambda^2$ so the structure of the peaks is more visible). Spectrum (a) shows a typical low-extinction O-rich spectrum where the set of peaks identified in Fig.~\ref{fig:example_spectra} are clearly seen. As we move around the crescent to spectrum (b) and onto spectrum (c) the set of O-rich peaks is still visible but the spectra are increasingly reddened. There are also less obvious variations in the shapes of the absorption features (which are modulated by the varying extinction) with the final peak at $\sim910\,\mathrm{nm}$ broader and boxier in spectrum (a) than in spectrum (b). This is a reflection of varying effective temperatures producing varying depths of the TiO$\delta(\Delta\nu=0)$ absorption feature. A similar sequence is seen moving up the spur from spectrum (d) which shows the three peaks of a C-rich spectrum seen in Fig.~\ref{fig:example_spectra} through to spectrum (e) which is distinctly more reddened but the peak structure is still visible. We attribute the relative narrowness of the C-rich spur to degenerate changes in both effective temperature and extinction rather than any change in the surface chemistry as the CN features are not strong functions of effective temperature. However, the broadness of the O-rich crescent is because both extinction and effective temperature (through the relative depths of TiO features) control the spectrum shape and can be distinguished. There are a couple of exceptions to these broad trends due to effective temperature and extinction variation which are apparent as small islands in Fig.~\ref{fig:umap}. The very small island (f) just off the C-rich spur has a median spectrum showing clear H$\alpha$ emission characteristic of a symbiotic system \protect\citep[this island contains the source identified by][]{Miszalski2013}. Mira variables also show strong Balmer series emission lines that vary with the pulsation phase but typically the line-to-continuum flux ratio is at most $\sim2$ \citep{Yao2017} whilst the symbiotic carbon star from \cite{Miszalski2013} has a line-to-continuum ratio of $\gg5$. It is likely Mira variable emission lines cannot be detected in BP/RP spectra. Finally, island (g) sits partway between the crescent and spur and its median spectrum shows evidence of both O-rich and C-rich behaviour. This is characteristic of an S-star that has intermediate chemistry C/O$\sim1$. We will discuss these stars further below.

\subsection{Model variants}
\begin{figure*}
    \centering
\includegraphics[width=\textwidth]{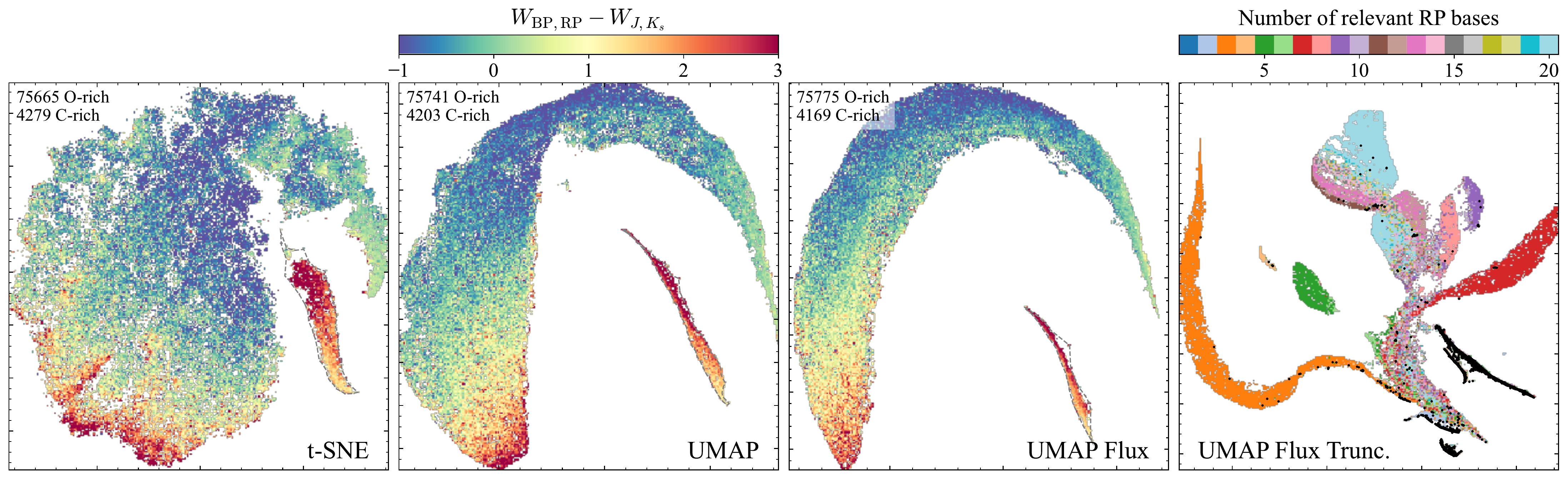}
    \caption{Variants of our unsupervised learning approach -- the left three panels show the t-SNE projection, the UMAP projection of the BP/RP coefficients (identical to Fig.~\ref{fig:umap}) and the UMAP projection of the binned spectra respectively, all coloured by the Wesenheit colour $W_\mathrm{RP,BP-RP}-W_{Ks,J-Ks}$ The grey contour shows the group identified by DBSCAN that we label as C-rich stars. The right panel shows the UMAP projection of the binned spectra computed from the truncated spectra coloured by the number of relevant RP bases. The black points are C-rich stars identified from the UMAP projection of the coefficients. Clearly, the stars have been separated more on the basis of their truncation order than on any intrinsic similarities in the data.}
    \label{fig:umap_var}
\end{figure*}
Before discussing in more detail the structure of the UMAP projection we briefly mention other approaches for separating the spectra. As a comparison, we have run some variants of our algorithm. First, we use t-SNE. t-SNE differs from UMAP primarily in the properties of the high-dimensional similarity distribution that are retained in the lower-dimensional projection. The cross-entropy used in UMAP preserves both a sense of local and global structure, whilst the t-SNE Kullback-Leibler divergence preserves only local structure (as discussed in the introduction). We use the open t-SNE implementation from \cite{openTSNE} on the full $2\times55-1$ dimensional space. A high perplexity (of $\sim100$) seems a good choice to maximise the separation between the two clusters. We show the results in Fig.~\ref{fig:umap_var} and see that there is a clear island that we can associate with C-rich stars. However, it is not as clearly separated from the bulk of the O-rich spectra. Furthermore, there is significant clumping within the O-rich region which we believe is more a reflection of the high perplexity choice than any sets of stars within the expected smooth continuum. Also, the S-star island is associated with the C-rich island whilst the more global metric preserving UMAP places it closer to the O-rich island.

\begin{figure*}
    \centering
\includegraphics[width=\textwidth]{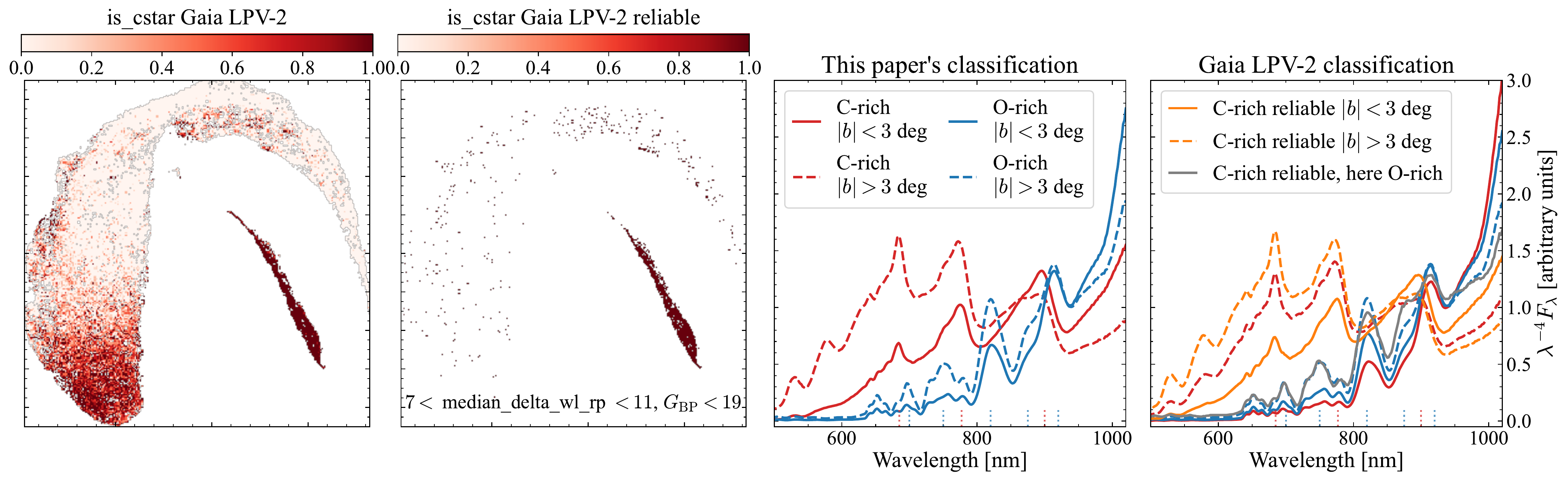}
    \caption{Comparison with the Gaia DR3 2nd LPV catalogue classifications. The left panel shows the UMAP projection of the high-amplitude sample coloured by the \texttt{is\_cstar} flag from the Gaia catalogue. The second panel restricts to the reliable flags with $7 < \texttt{median\_delta\_wl\_rp} < 11$ and $G_\mathrm{BP} < 19$. In the right two panels, we show the median normalized BP/RP spectra classified using this paper's methodology (third panel) and the Gaia DR3 methodology (fourth panel). We split by those classified as C-rich (red) and O-rich (blue) at high ($|b|>3\,\mathrm{deg}$, dashed) and low ($|b|<3\,\mathrm{deg}$, solid) latitudes respectively. In the fourth panel, the orange lines show the median of the reliable C-rich classifications from Gaia. Finally, the grey line in the right panel shows the median of the spectra classified here as O-rich but as reliable C-rich in the Gaia DR3 release. From this, it is evident that the reportedly reliable classifications from Gaia DR3 are on the whole robust for separating O-rich and C-rich, but the suspected unreliable low-latitude C-rich classifications are primarily O-rich stars.}
    \label{fig:gaia_panel}
\end{figure*}

We further experiment by using as inputs to UMAP the externally-calibrated spectra on a wavelength grid ($336$ to $1020\,\mathrm{nm}$ with $2\,\mathrm{nm}$ spacing). We first normalize the spectra by dividing them by the sum of the binned flux values before performing the same procedure as applied to the coefficients. We show the results in the third panel of Fig.~\ref{fig:umap_var}. We have found in general there is a very similar degree of separation using both the coefficient and the flux space. The dimensionality of the coefficient space is smaller (although our wavelength grid choice here is arbitrary) and the dimensions are anticipated to be less correlated than for the sampled flux space so we prefer the coefficient space.

Finally, we investigate truncating the BP/RP coefficients using the \texttt{xp\_n\_relevant\_bases}. For each star, we set all higher-order coefficients to zero and construct the normalized externally-calibrated spectra on the previously mentioned wavelength grid. Running UMAP on this set produces the right panel of Fig.~\ref{fig:umap_var}. This clearly has significantly more structure than when using the full spectra without truncation and colouring by truncation order in RP makes it clear that each cluster is linked to a specific truncation order. Furthermore, the stars identified as C-rich in the UMAP projection of the coefficients are split and not clearly identifiable as an association. It, therefore, seems, at least in this case, that truncation is significantly deteriorating the results. Our case may be special as we are considering very red (both intrinsically and due to interstellar dust) and variable objects. There is clearly significant information in the higher-order terms that is being neglected by a hard cut. As we will see, the basis expansion also appears sub-optimal for the C-rich stars as the chi-squared/standard deviation of the fit for these stars is significantly larger than for the O-rich. 

\subsection{Comparison with Gaia DR3 classifications}
In the Gaia DR3 LPV candidate catalogue \citep{Lebzelter2022}, there is already a flag for whether a source is O-rich or C-rich on the basis of the BP/RP spectra. This identification is based upon the separation of the most prominent peaks in the (internally-calibrated) RP spectra with C-rich spectra having more separated peaks than O-rich stars. As already noted by \cite{Lebzelter2022}, the Gaia DR3 LPV \texttt{is\_cstar} flag can produce unreliable results. \cite{Lebzelter2022} advises only using the classification if $7 < \texttt{median\_delta\_wl\_rp} < 11$ and $G_\mathrm{BP} < 19$. Without this cut, there is a high number of C-rich LPVs in the Galactic midplane and in particular in the Galactic bulge, somewhat in tension with the previous work discussed in the introduction. It appears that in the presence of significant interstellar dust the identification of the peaks fails possibly because the spectrum is so skewed to the red that the red edge of the RP bandpass is being identified as a peak associated with an absorption feature. In the lower panels of Fig.~\ref{fig:gaia_panel}, we show the UMAP diagram coloured by the \texttt{is\_cstar} flag both for the entire set and restricting to the reliably flagged stars. This demonstrates that the main cluster we have identified as C-rich is populated by stars reliably flagged as C-rich through the method of \cite{Lebzelter2022}. However, there is also a clear overabundance of stars with non-zero \texttt{is\_cstar} in the high-extinction end of the O-rich cluster. Even of those stars reliably flagged as C-rich, there are many that reside in the main O-rich cluster.

In the right panels of Fig.~\ref{fig:gaia_panel}, we show the median of the normalized spectra for those stars we identify as O or C-rich split between a low- and high-latitude sample at $|b|=3\,\mathrm{deg}$. We see the common features in both the low and high-latitude samples for each set. In the rightmost panel, we display similar using the \texttt{is\_cstar} flag (not restricting to the reliable subset). Clearly, the low-latitude C-rich sample very closely resembles the low-latitude O-rich sample. Restricting to the reliable \texttt{is\_cstar} subset, we see the high- and low-latitude C-rich samples have the characteristic C-rich features. Finally, we show the median spectrum of the reliable \texttt{is\_cstar} stars that we have identified as O-rich. Again, this has evidence of the O-rich peak structure so we believe our assignment is more robust. Performing the same tests but separating the sample instead on $A_0$ from the Gaia DR3 optimized total galactic extinction map yields near-identical conclusions. In this case, however, even the reliable C-rich Gaia LPV-2 classifications for the high-extinction sample are clearly predominantly O-rich. In conclusion, we have found that our method offers an improvement over the Gaia DR3 LPV catalogue, especially for the reddest most extincted sources in Gaia. This is perhaps not surprising as we have used all the information from the spectra.

In the Gaia DR3 data release, there is a table of `golden carbon stars' that have been selected from an initial list classified from the BP/RP spectra using a random forest classifier trained on synthetic data and Gaia data for known Galactic carbon stars \citep{Creevey2022}. The initial list of $386,936$ candidates was filtered using the strength of the right two peaks marked on the C-rich spectrum of Fig.~\ref{fig:example_spectra} to reduce the sample down to $15,740$ `bona fide' carbon stars. $13,513$ of the golden carbon stars are classified as LPVs in Gaia DR3 of which we classify $13,239$ as C-rich using the classification of the full set in Appendix~\ref{appendix::full}. $61$ are classified as O-rich and $213$ do not have BP/RP spectra in the Gaia DR3 data release. We have a total of $23,737$ C-rich classifications based on the BP/RP spectra. For the high-amplitude subset described in the body of this paper, there are $1835$ in the golden carbon star list of which only $4$ are classified as O-rich by our algorithm. We find a total of $4203$ C-rich LPV stars with BP/RP spectra. 

\subsection{The structure of the UMAP projection}

\begin{figure*}
    \centering
\includegraphics[width=\textwidth]{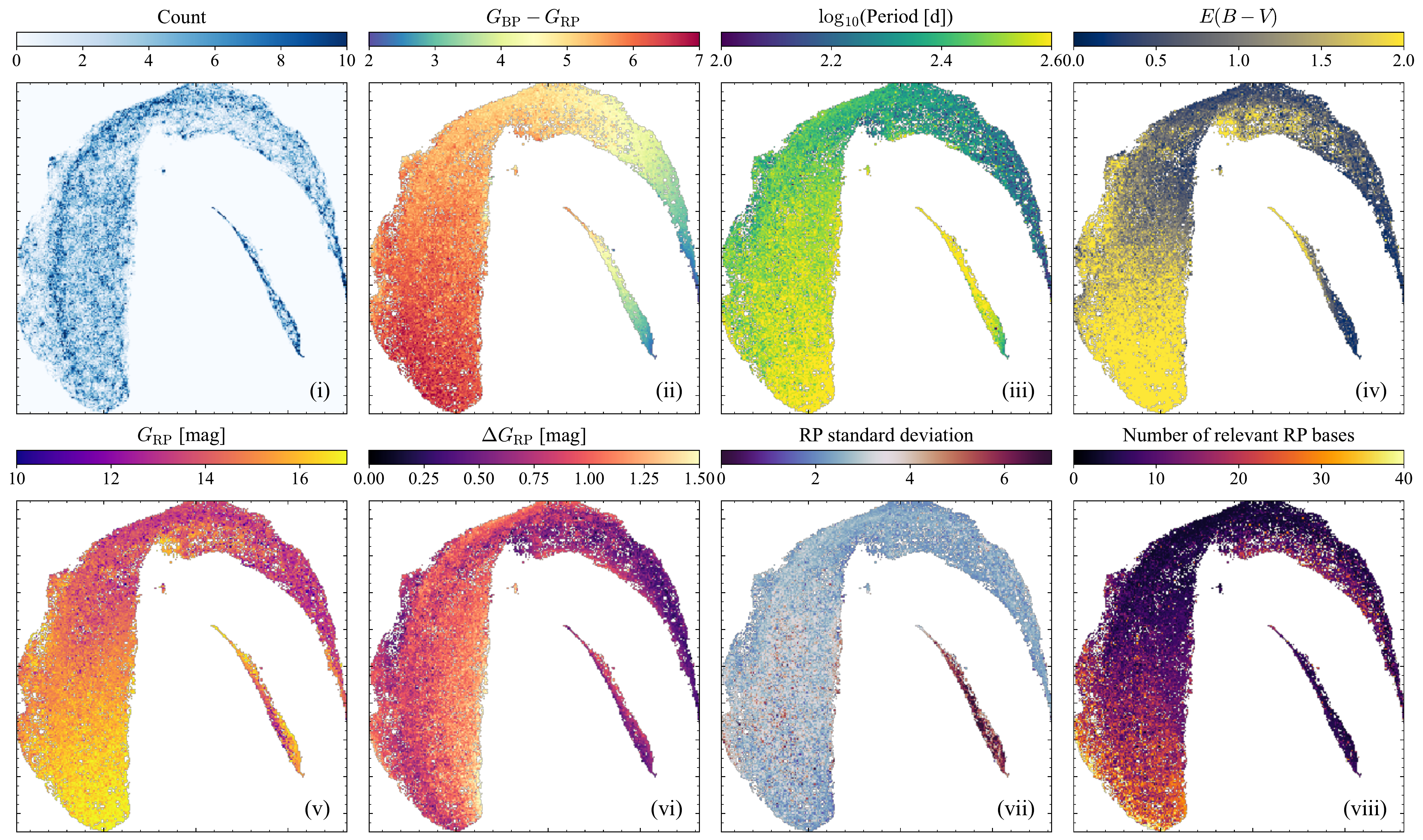}
    \caption{Two-dimensional UMAP projections coloured by the means of different binned quantities. We show the UMAP projection of the coefficients coloured by (i) the count per bin, (ii) the $G_\mathrm{BP}-G_\mathrm{RP}$ colour, (iii) the logarithm of the period, (iv) the \protect\cite{SFD} interstellar extinction, (v) the $G_\mathrm{RP}$ magnitude, (vi) the $G_\mathrm{RP}$ amplitude, (vii) the standard deviation of the RP spectrum fit and (viii) the number of relevant RP basis functions.}
    \label{fig:multi_panel}
\end{figure*}

We now investigate further the structure of the UMAP projection shown in Fig.~\ref{fig:umap}, in particular focusing on the internal structure of the crescent and spur. Figure~\ref{fig:multi_panel} shows the UMAP projection coloured by various properties. From the UMAP diagrams, it is clear that both the O-rich and C-rich stars form approximately one-dimensional sequences with the O-rich sequence having a more significant width perpendicular to this. However, even along the O-rich crescent, there is a ridge line. The one-dimensional sequences are largely arising due to broad $(G_\mathrm{BP}-G_\mathrm{RP})$ variation as evidenced by panel (ii). This can arise either through effective temperature or extinction variation. For general stars, it is quite difficult to separate effective temperature and extinction variation using broadband photometry as the extinction vector typically lies near parallel to the stellar sequence. However, absorption line structure from spectroscopy is sensitive to the effective temperature of the stars and so allows for extinction-independent measurements. As seen in panel (iv), the interstellar $E(B-V)$ from \cite{SFD} approximately increases along the crescent. Using $A_0$ from the Gaia DR3 optimized total galactic extinction map produces near identical trends for the $75\percent$ of our sample that fall in the on-sky region covered by the map. This trend with extinction suggests the direction perpendicular to the extinction gradient is due to detected effective temperature variation. This is somewhat evidenced by the amplitude variation, $\Delta G_\mathrm{RP}$, across the sequence shown in panel (vi). The period colouring in panel (iii) shows that period is also increasing around the sequence. This could be arising from intrinsic variations with longer-period cooler stars displaying distinct spectral features, perhaps arising from more circumstellar extinction through higher mass loss, or it could be because longer-period stars are associated with younger populations and so more confined to the higher extinction midplane. As already described, the relative abundance of molecular species gives a probe of effective temperature. As CN forms deeper in the atmosphere and is largely independent of the surface temperature for these stars, there is little effective temperature variation in the observed BP/RP spectra for C-rich stars and they lie along a narrow sequence in the UMAP diagram due almost entirely to extinction. On the other hand, in O-rich stars, it is expected that the abundance of TiO is a strong function of effective temperature \citep{Reid2002}. We define the relative depth of the TiO features as
\begin{equation}
\delta\mathrm{TiO} = \frac{\mathrm{TiO}\epsilon(\Delta\nu=-1)-\mathrm{TiO}\epsilon(\Delta\nu=0)}{\mathrm{TiO}\epsilon(\Delta\nu=0)-\mathrm{TiO}\delta(\Delta\nu=-1)},
\label{eqn::tio}
\end{equation}
where
\begin{equation}
\mathrm{TiO}j = \int_{\lambda_j-\delta\lambda}^{\lambda_j+\delta\lambda} \mathrm{d}\lambda\,F_\lambda,
\end{equation}
and $(\lambda_{\delta(\Delta\nu=-1)},\lambda_{\epsilon(\Delta\nu=0)},\lambda_{\epsilon(\Delta\nu=-1)})=(779,853,940)\,\mathrm{nm}$ are the locations of three prominent TiO bands as identified in Fig.~\ref{fig:example_spectra}. We choose $\delta\lambda = 15\,\mathrm{nm}$. $\delta\mathrm{TiO}$ is defined to be less sensitive to the broad spectrum shape which is sensitive to both effective temperature and extinction. Colouring the UMAP diagram by this feature as shown in Fig.~\ref{fig:s_star}, it is clear that the TiO band depth (or effective temperature) is varying across the O-rich UMAP sequence. This is particularly evident in the top half of the spur. In figure 4 of \cite{Lancon2000}, in warmer stars the depths of $\mathrm{TiO}\epsilon(\Delta\nu=0)$ and $\mathrm{TiO}\delta(\Delta\nu=-1)$ are comparable but $\mathrm{TiO}\epsilon(\Delta\nu=-1)$ is weak. For the cooler stars, the $\mathrm{TiO}\epsilon(\Delta\nu=-1)$ depth increases to be similar to the relative depths of $\mathrm{TiO}\epsilon(\Delta\nu=0)$ and $\mathrm{TiO}\delta(\Delta\nu=-1)$. Therefore, $\delta\mathrm{TiO}$ decreases as the star gets cooler.

Panel (vii) of Fig.~\ref{fig:multi_panel} displays the standard deviation of the Gaia RP spectrum fit, i.e. the chi-squared per degree of freedom, as provided in the Gaia DR3 catalogue. Interestingly this displays a very clear separation between O-rich and C-rich sources with the C-rich sources having much poorer fits. This does not appear to be linked to their typically slightly fainter magnitudes (panel (v) shows there are many O-rich stars of similar magnitudes with similar quality fits) nor is it linked to their amplitudes as C-rich stars actually typically have lower amplitudes in the visual bands than the O-rich stars. Our explanation is that the RP basis function choice is optimized for a set of calibrators and C-rich stars are probably a minority population in this set (if present at all). As the C-rich spectral features are very distinct from the more typical O-rich stars, the RP basis functions don't completely capture the behaviour of C-rich stars. It is interesting then that more generically the RP standard deviation could be utilised to identify C-rich stars (although note that some O-rich stars can also have poorly fitting solutions for other reasons). Despite the poorer fits of the C-rich stars, panel (viii) shows that they have a low number of relevant RP bases. This is likely again because the ordering of the bases has been based upon typical calibrator stars whilst C-rich stars probably have insignificant information in intermediate terms (that may capture the molecular features in an O-rich spectrum) and more significant information in the higher order terms. Note as well that higher extinction O-rich stars require more terms to capture their behaviour.

\subsubsection{S-stars and comparison with literature classifications}
\begin{figure}
    \centering
\includegraphics[width=\columnwidth]{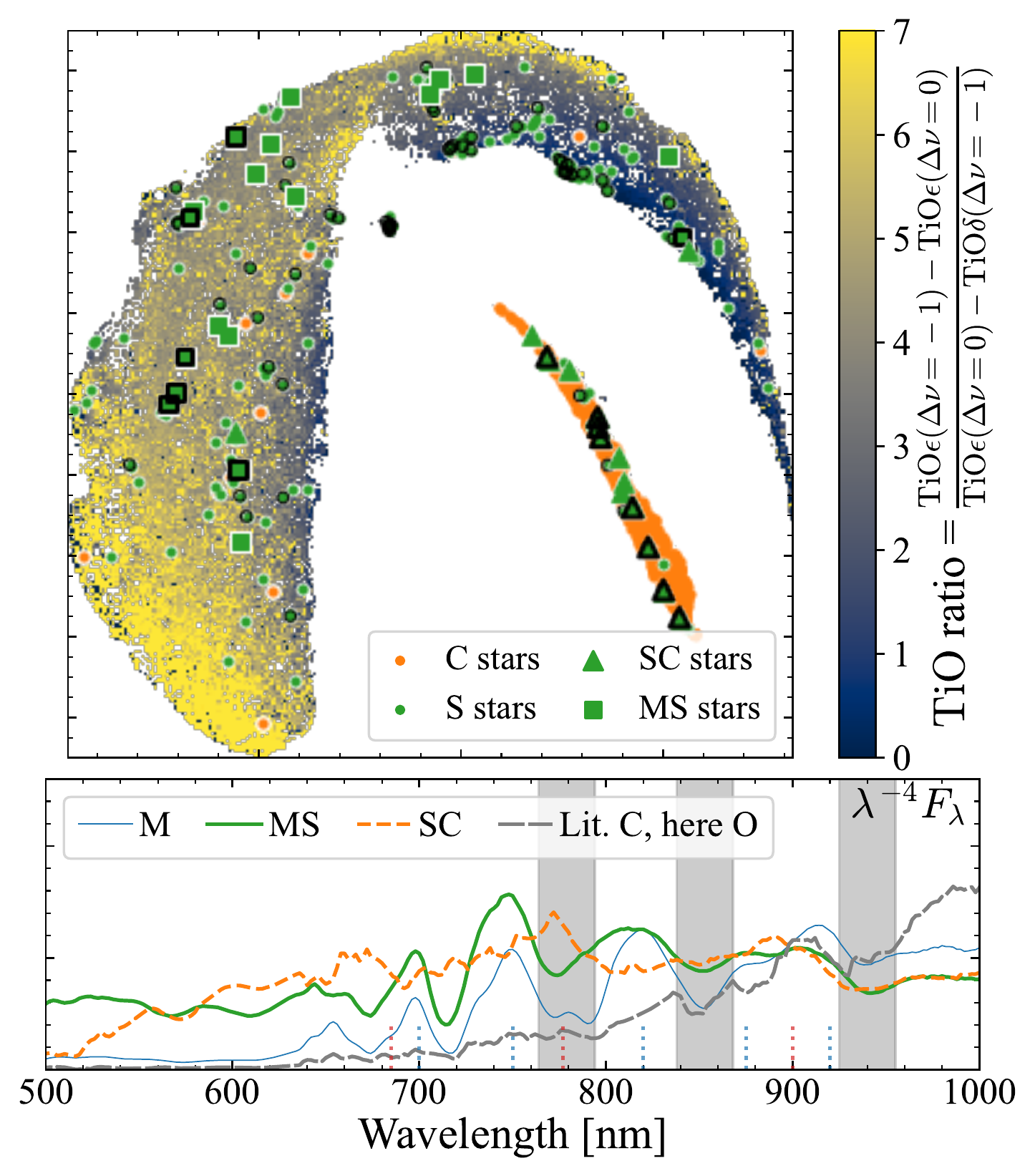}
    \caption{Comparison with literature identifications of C-stars (orange) and S-stars (green). The top panel shows the UMAP projection coloured by the TiO ratio constructed in equation~\eqref{eqn::tio}. The SC stars are triangles and the MS stars are squares. Any stars with emission lines are outlined in black. The lower panel shows the median spectra of O-rich stars (M, thin blue), MS stars (green thicker), SC stars (orange short-dash) and those stars classified as C-rich on SIMBAD but classified as O-rich here (grey long-dashed).}
    \label{fig:s_star}
\end{figure}

S-stars have intermediate C/O $\sim1$ and subsequently chemistry that shares characteristics with both O-rich and C-rich stars. One identifying feature is the presence of ZrO molecular features. We already observed from Fig.~\ref{fig:umap} that a small clump of stars near the O-rich crescent had median BP/RP spectra indicative of S-stars. We have taken all S-stars from SIMBAD \citep{Simbad} and matched $240$ to our sample. We show their locations on the UMAP plane in Fig.~\ref{fig:s_star}. We also separate out MS type (those S-stars with more O-rich chemistry) and SC type \citep[those with more C-rich chemistry, see][for a clear illustration of the progression]{Yao2017}. The SC stars all lie on the C-rich spur whilst the MS stars lie in the O-rich crescent. Although they are distributed across the entire crescent, there are a number of overdensities, particularly on the underside of the crescent, and also crucially on the small island identified in Fig.~\ref{fig:umap}. We reason that the overdensity of S-stars along the underside of the crescent is due to ZrO in these stars which lies at the same location in the spectrum as TiO$\epsilon(\Delta\nu=-1)$. \cite{VanEck2017} show a series of spectra with increasing C/O where the TiO features weaken but the absorption at $940\,\mathrm{nm}$ stays quite constant due to the increasing contribution of ZrO. We have also indicated those stars classified in SIMBAD as emission line objects (presumed mostly from the identification of H$\alpha$ and other Balmer lines, but this is a heterogeneous set) but these appear to be indistinguishable from the bulk of the stars.

Furthermore, we have obtained $10,000$ C stars from SIMBAD \citep{Simbad} and found $805$ matches in our dataset. These are also displayed in Fig.~\ref{fig:s_star}. The majority live along the C-rich spur with a few in the O-rich crescent. The median spectrum of the objects `misclassified' by us as O-rich is shown in the lower panel of Fig.~\ref{fig:s_star}. It is not clear what the exact nature of these misclassified stars is but one can clearly see the TiO absorption at $\sim850\,\mathrm{nm}$ so we are inclined to classify them as O-rich.

\subsection{Supervised classification}\label{sec:supervisedClassification}

\begin{figure*}
\centering
    \includegraphics[width=\textwidth]{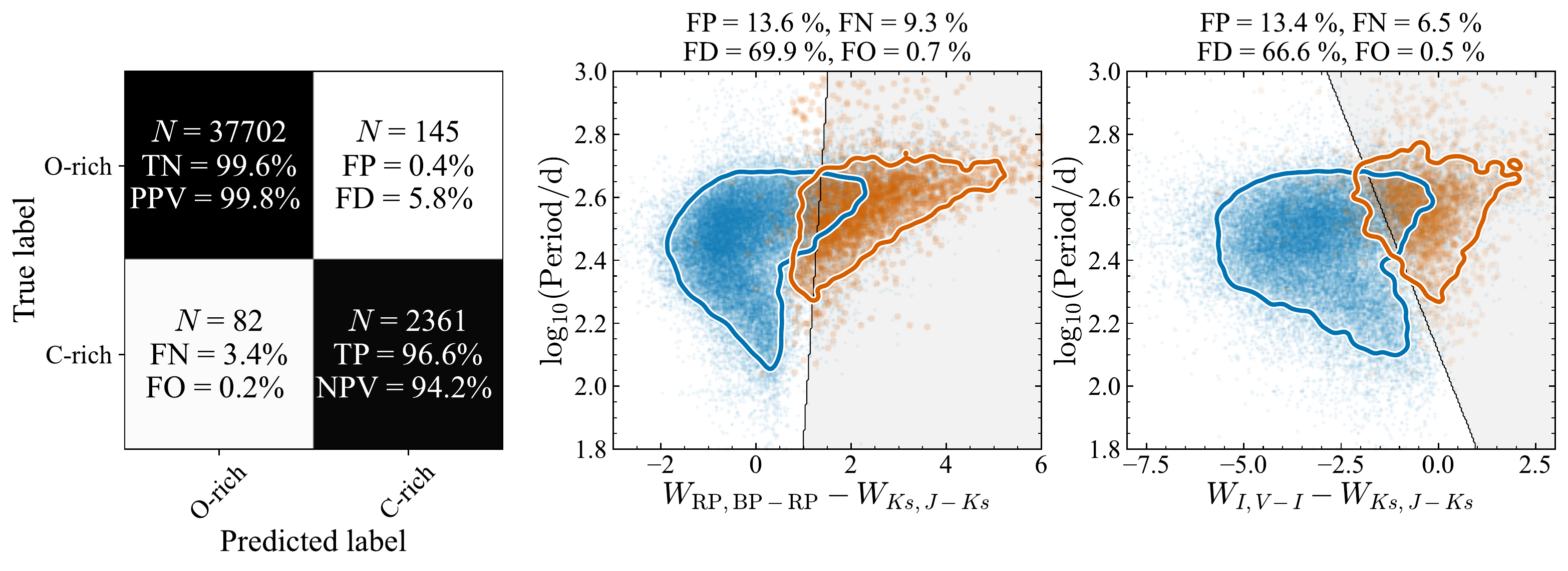}
    \caption{Supervised classification schemes: the left panel shows the confusion matrix for the application of XGBoost to the classification of O/C-rich stars from photometric colours, periods and amplitude. We report the number and other statistics described in the text. The central panel shows two other optical and infra-red photometric spaces for separating O/C-rich stars. The blue and orange clouds show the O-rich and C-rich spectroscopic classifications with the contours containing $80\percent$ of each dataset. The grey line shows the best linear separation between the two classes.}
    \label{fig:supervised}
\end{figure*}

As BP/RP spectra are only available for stars with $G<17.65$, utilising the BP/RP classifications alone would remove many highly-extincted stars. From the work of \cite{Lebzelter2018} it is clear broadband optical and infrared photometry can be used effectively to separate O-rich and C-rich sources. We use the previous classifications to train a gradient-boosted random forest classifier \citep[XGBoost]{xgboost}. Due to the imbalance of the dataset, we use weights inversely proportional to the number of each class in the dataset. We have found the best performance is obtained using ($J-K_s$, $G_\mathrm{BP}-G_\mathrm{RP}$, $G-G_\mathrm{RP}$, Period, $\Delta G_\mathrm{RP}$), where $\Delta G_\mathrm{RP}$ is computed from \texttt{std\_dev\_mag\_rp} (see Appendix~\ref{section::amplitude_comparison}). We limit ourselves to considering stars with high-quality 2MASS photometry (\texttt{ph\_qual}=`A') and a low fraction of blended/contaminated BP and RP observations (\texttt{phot\_bp/rp\_n\_contaminated/blended\_transits}
$/$
\texttt{phot\_bp/rp\_n\_obs}$\leq0.1$) \citep{Riello2021}. The resulting feature importance is $(
0.19, 0.52, 0.13, 0.14, 0.03
)$. We also store the classification probabilities from XGBoost. We give the resulting confusion matrix in the left panel of Fig.~\ref{fig:supervised} where we quote the total numbers of correct and incorrect classifications, the true positive, true negative, false positive and negative rates (all normalized by the number of true classifications) and the positive predictive value (PPV, positive predicted as positive), the negative predictive value (NPV, negative predicted as negative), the false discovery (FD) rate (negative predicted as positive) and the false omission (FO) rate (positive predicted as negative, all normalized by the number of predictions made). Here `positive' is a C-rich classification. For the identification of C-rich stars, the false discovery rate (related to the purity of the sample) of the C-rich predictions is the most important. Here we find $5.8\percent$. We further only lose FN=$3.4\percent$ of genuine C-rich stars so the completeness is also high. As the number of O-rich stars overwhelms C-rich stars, the purity of the O-rich sample (PPV) is very high ($99.8\percent$). The classifier metrics are weak functions of $G$ making their extrapolation to the fainter stars without BP/RP spectra valid.

We further inspect previously employed classification schemes based on optical and near-infrared data. As already evidenced in Fig.~\ref{fig:umap}, the Wesenheit colour-colour indicator from \cite{Lebzelter2018} performs well to separate O-rich and C-rich even for non-LMC stars. We display the projection in the central panel of Fig.~\ref{fig:supervised}. We perform a simple linear support vector machine classification in the $W_\mathrm{RP,BP-RP}-W_{Ks,J-Ks}$ vs. period space for the same sample of data used in the XGBoost models. The projection does perform well for low periods but for high extinction, the O-rich stars enter into the low-extinction C-rich region leading to high false discovery rates/contamination in any C-rich sample (FD is around $95\percent$ for $E(B-V)>3$). However, this space is appropriate for removing C-rich stars from an O-rich sample. Restricting to $|b|\gtrsim5\,\mathrm{deg}$ largely removes the  highly-reddened sources. Although we do not have access to the OGLE photometry for the majority of our sample, as the BP/RP spectra cover the entire range of the OGLE $V$ and $I$ bands we can use them to simulate what OGLE would see \citep{DeAngeli2022}. We use the filters from the SVO filter service \citep{SVO1,SVO2} and sum the BP/RP spectra on the wavelength grid reported for the filter.  A cross-check for those stars with measured OGLE photometry demonstrates this procedure performs reasonably well but there are large uncertainties, particularly for faint $V$. We display the resulting distribution in the right panel of Fig.~\ref{fig:supervised}. It largely resembles the $W_\mathrm{RP,BP-RP}-W_{Ks,J-Ks}$ vs. period projection but slightly rotated. For this reason, a linear support vector classifier performs similarly and suffers the same issue with highly-reddened stars.

\section{Validation of the classification scheme}\label{sec::validation}

\begin{figure*}
\centering
    \includegraphics[width=\textwidth]{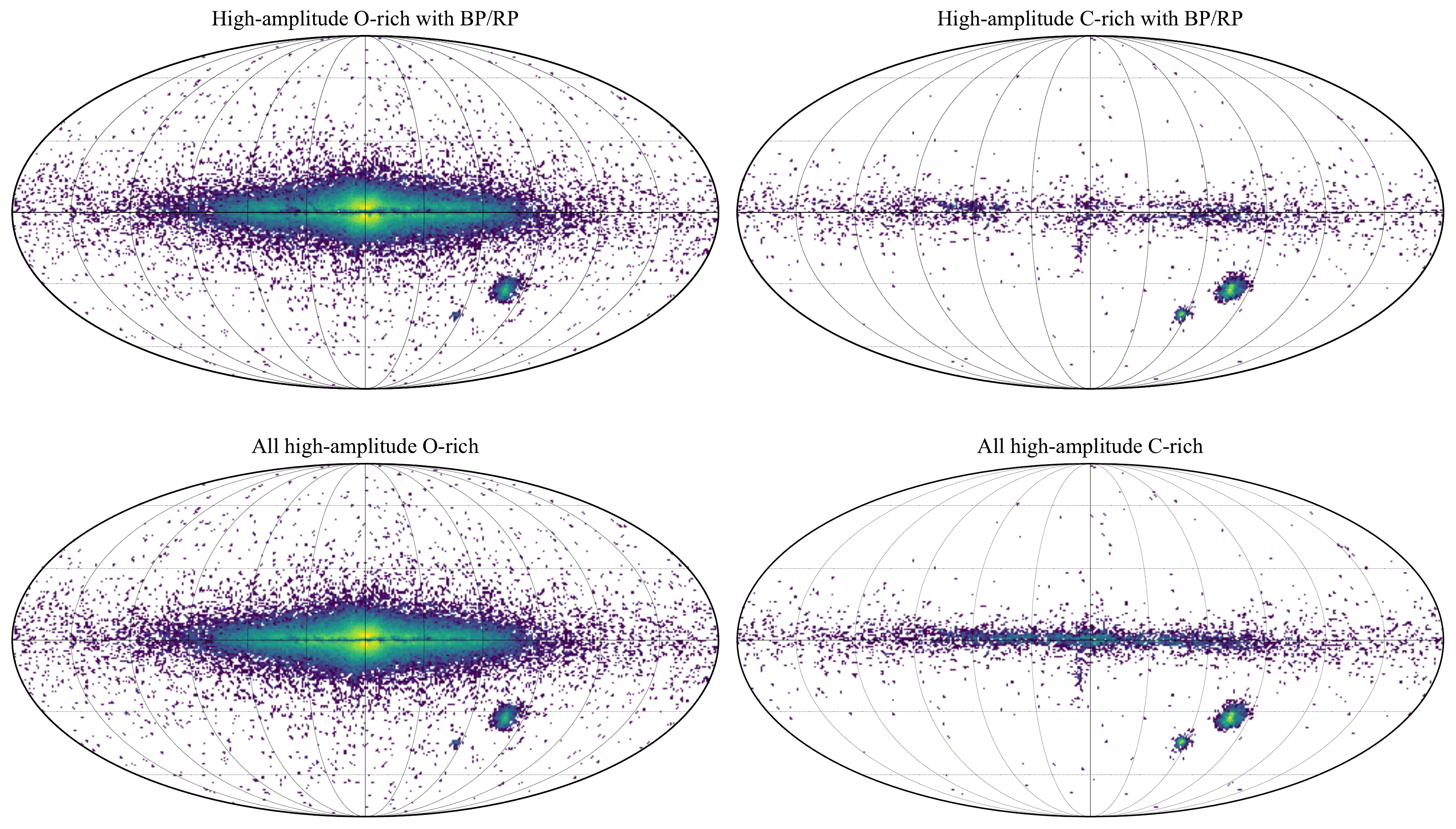}
    \caption{On-sky distribution in Galactic coordinates of the unsupervised classifications based on BP/RP spectra (top row) and the supervised classifications using photometry. The left panels are O-rich classification whilst the right are C-rich. }
    \label{fig:onsky}
\end{figure*}

\begin{figure*}
    \centering
    \includegraphics[width=\textwidth]{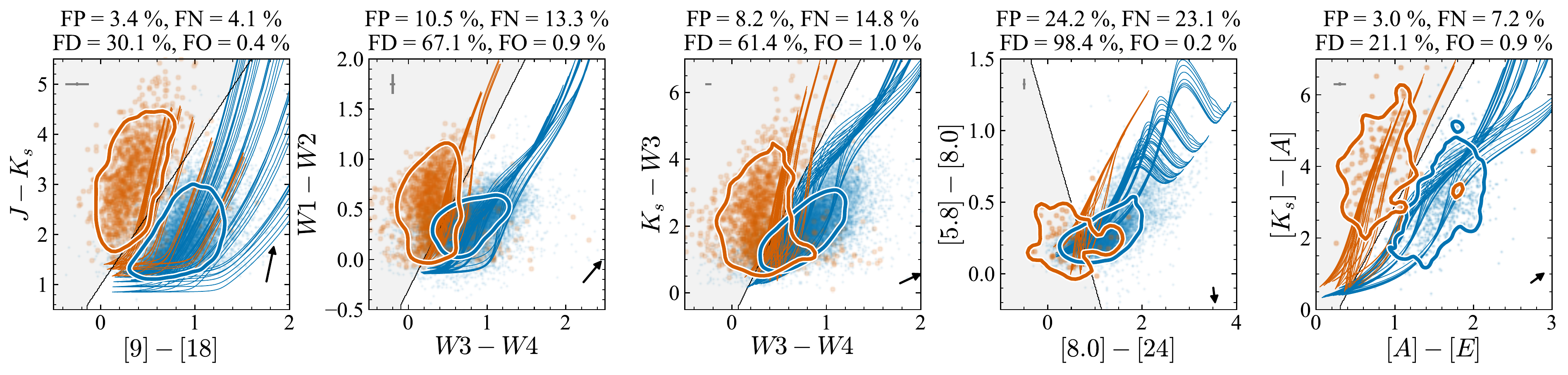}
    \caption{Infrared colour-colour diagrams with the classified O-rich (blue) and C-rich (orange) high-amplitude LPVs. The contour contains $80\percent$ of each set. The lines are sequences of dusty O-rich and C-rich models. The grey region shows the best linear support vector classifier for separating O-rich and C-rich. We report the false positive (FP, percentage of O-rich stars classified as C-rich), false negative (FN, percentage of C-rich stars classified as O-rich), false discovery (FD, percentage of stars classified as C-rich that are O-rich) and false omission (FO, percentage of stars classified as O-rich that are C-rich) above each plot. The small arrow is the extinction vector direction using the results from \protect\citet[][the photometry is not dereddened]{Fritz2011} and the grey errorbar is the median uncertainty in the photometry.}
    \label{fig:infrared_diagrams}
\end{figure*}

In Fig.~\ref{fig:onsky} we display the on-sky distributions of the O-rich and C-rich long-period variable stars based on our unsupervised and supervised schemes. We see that in agreement with previous works the Galaxy is dominated by O-rich variables and the C-rich variables are biased more towards the outer Galactic disc. We also note the comparative excess of C-rich variables in the Small Magellanic Cloud and the Sagittarius dwarf spheroidal galaxy. We also notice the Galactic bulge contains some C-rich stars -- we will return to this later. Our classification seems to agree with previous work indicating the bulk of long-period variables in the Milky Way are O-rich and the relative fraction of C-rich variables increases in the outer disc \citep{Blanco1984,Ishihara2011}.

We further validate our classification procedure by comparison to previously employed schemes based on colour--colour infrared photometry diagrams.
We perform cross-matches of the sample with BP/RP spectra classifications to GLIMPSE, MSX, AKARI, WISE and 2MASS (using a $1\,\mathrm{arcsec}$ crossmatch radius except for AKARI where we use $3$). We apply the $4$-band zero-point corrections to the All-WISE data listed at \url{https://wise2.ipac.caltech.edu/docs/release/neowise/expsup/sec2_1civa.html}. In Fig.~\ref{fig:infrared_diagrams} we show five commonly employed colour--colour diagrams and display our classified objects. We further overlay the set of dusty AGB models from \cite{Sanders2022}. The AKARI projection has been advocated by \cite{Ishihara2011} and \cite{Matsunaga2017}, the WISE diagram by \cite{Lian2014}, the WISE/2MASS diagram \cite{SuhHong2017}, the GLIMPSE diagram by \cite{Groenewegen2018} and the MSX/2MASS diagram by \cite{Lewis2020a,Lewis2020b}. Clearly in all but the GLIMPSE diagram, our classification produces two distinct clusters of points. In the GLIMPSE diagram, there is a low number of C-rich sources and also the extinction acts to make stars bluer in $([5.8]-[8.0])$. We run a linear support vector machine classifier in each colour--colour space, balancing each class using weights inversely proportional to their number in the dataset. Above each panel of Fig.~\ref{fig:infrared_diagrams} we give the false positive, false negative, false discovery and false omission percentages of the C-rich classifications i.e. the fraction of `true' O-rich stars classified as C-rich, the fraction of `true' C-rich stars classified as O-rich, the fraction of C-rich classifications that are `truly' O-rich and the fraction of O-rich classifications that are `truly' C-rich. Both the AKARI/2MASS $([9]-[18])$ vs. $(J-K_s)$ space \citep[advocated by][]{Matsunaga2017} and the MSX/2MASS $([A]-[E])$ vs. $([K_s]-[A])$ space \citep[advocated by][]{Lewis2020a,Lewis2020b} produce good separations of the populations with false positive rates for O-rich and C-rich classification of around $5\percent$. The other diagrams are typically poorer due to the overlap in O-rich and C-rich stars in the bluer parts of the diagrams suggesting the differing circumstellar dust is the primary driver for the separation in these diagrams and when it is absent, there is limited photometric difference between the populations. The reported statistics for each colour--colour diagram do not reveal the true efficacy of each colour--colour diagram as they are biased towards those Mira variables that are optically detected in Gaia. This naturally misses very red sources possibly highly embedded in circumstellar dust. Such sources are preferentially C-rich, so we would typically expect more C-rich sources from an infrared catalogue. This suggests e.g. the false positive rate for the O-rich classification that we report is an optimistic (under-) estimate. However, from the models, it is evident that the redder sources are more easily distinguishable suggesting that even with redder C-rich sources in our sample, the false positive rate for the O-rich classification will not change significantly.

\section{Potential C-rich bar-bulge members}\label{sec::bulge}

\begin{figure*}
    \centering
    \includegraphics[width=\textwidth]{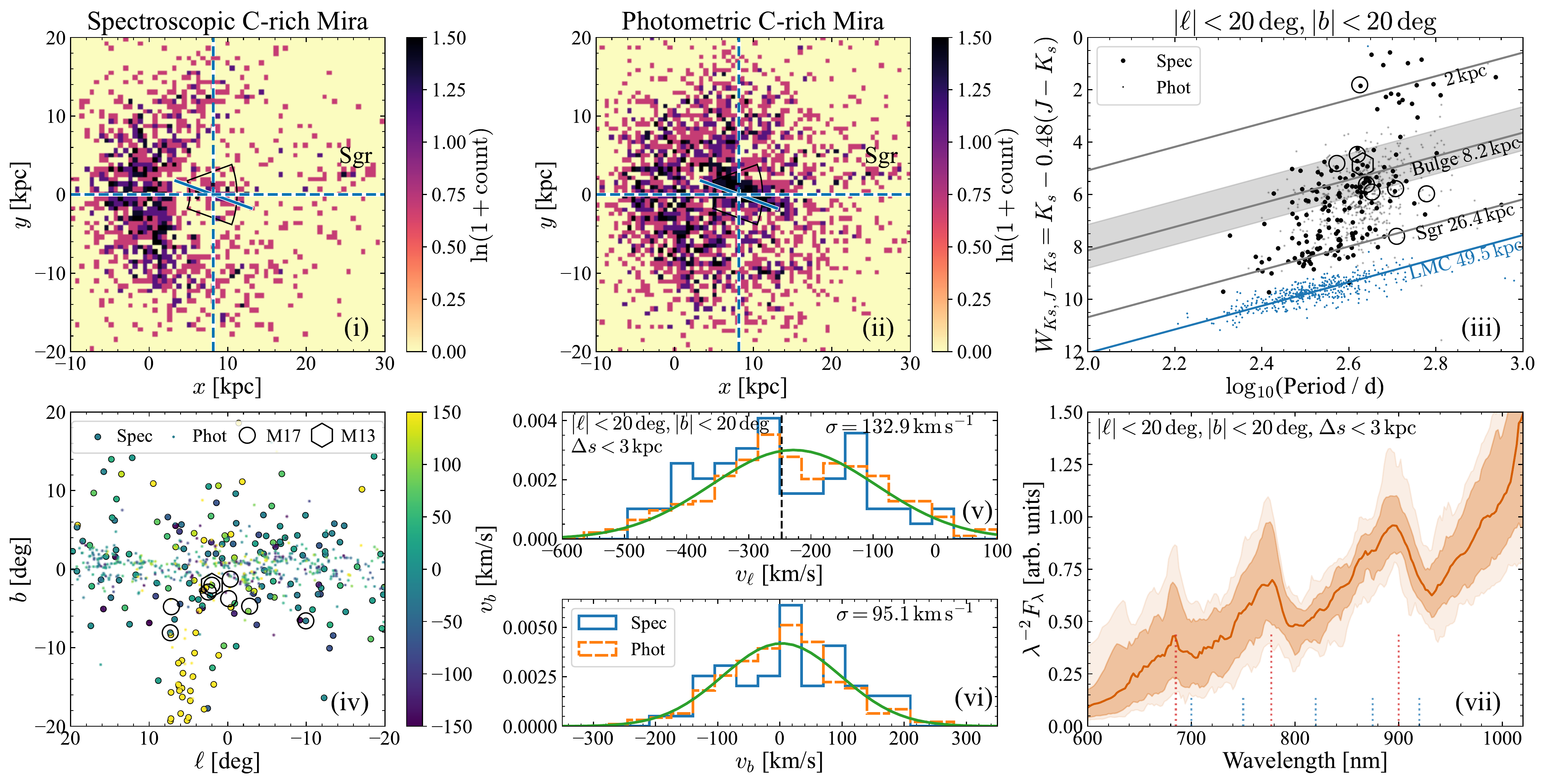}
    \caption{C-rich Mira variables within the Galactic bulge -- panels (i) and (ii) show the view from the Galactic North Pole of those high-amplitude ($>0.6\,\mathrm{mag}$) stars (i) spectroscopically and  (ii) photometrically classified as C-rich. The vertical line is at the Galactic centre distance and the small tilted line is at an angle of $20\,\mathrm{deg}$. Panel (iii) shows the Wesenheit magnitude $W_{Ks,J-Ks}=K_s-0.48(J-K_s)$ against period for spectroscopically identified C-rich stars within $10\,\mathrm{deg}$ of the LMC (blue dots) and then those within $|\ell|<20\,\mathrm{deg}$ and $|b|<20\,\mathrm{deg}$ both spectroscopically (large black) and photometrically (small black) identified. The hexagon is the symbiotic C-rich star from \protect\cite{Miszalski2013} and the circles are the C-rich Mira variables identified by \protect\citet[][using their mean photometry]{Matsunaga2017}. Panel (iv) shows stars in this region coloured by Galactic latitude proper motion with the \protect\cite{Miszalski2013} and \protect\cite{Matsunaga2017} stars also plotted. The histograms in panels (v) and (vi) show the transverse velocity distributions of stars in this region within $3\,\mathrm{kpc}$ of the Galactic centre (dashed is photometric identifications) with the best-fitting Gaussians to the photometric identifications in green. Panel (vii) gives the median BP/RP spectrum (with $\pm1,2\sigma$ brackets) for the spectroscopic C-rich bulge identifications.}
    \label{fig:crich}
\end{figure*}
We close this work by addressing some of the questions raised in the introduction, namely how many C-rich stars are there in the Galactic bulge and what is their nature. Fig.~\ref{fig:onsky} has already demonstrated that there is a low number of stars classified as C-rich from the BP/RP spectra. We first restrict to those stars with semi-amplitudes between $0.6$ and $2$ to remove any semi-regular variables and any spurious high amplitudes due to aliasing (see Appendix~\ref{section::amplitude_comparison}). We identify reliable spectroscopic C-rich as those lying on the C-rich spur from Fig.~\ref{fig:umap} and that have supervised cross-validated classification probabilities of being C-rich of $>0.9$. We remove potential young stellar object contaminants ensuring no stars have $G-5\log_{10}(100/(\varpi-3\sigma_\varpi))>2.5(G_\mathrm{BP}-G_\mathrm{RP})-5$ and also restrict to stars with Gaia DR3 classification probabilities $>0.5$ or those classified as `SYST'. For this subset, we have generated $100$ samples from the BP/RP coefficient covariance matrix and run them through the unsupervised classifier. If any of the per-star samples are classified as O-rich we remove the star from the sample. For the photometric C-rich candidates, we ensure similar criteria but also ensure any star isn't spectroscopically classified as O-rich. In this way, we end up with $2018$ and $2687$ spectroscopically and photometrically classified C-rich Mira variables respectively across the entire sky.

We display these samples in the top left panels of Fig.~\ref{fig:crich} as viewed from the Galactic North Pole. We have assigned approximate distances to the stars using the Wesenheit $W_{Ks,J-Ks}=K_s-0.48(J-K_s)$ vs. period relation for those stars within $10\,\mathrm{deg}$ of the LMC as shown in the top right panel of Fig.~\ref{fig:crich}. The extinction coefficient is appropriate for the Galactic bulge \citep{Nishiyama2009,Fritz2011,AlonsoGarcia2017,Sanders2022_Ext}. Note the relatively tight relation followed by the LMC stars giving confidence in our amplitude cut for isolating only those stars on the Mira variable sequence \citep{Wood2000}. We have fitted the linear relation $W_{Ks,J-Ks}=-4.5(\log_{10}(P/\mathrm{d})-2.3)+10.7$ by-eye to these stars and use the distance modulus of $18.477$ for the LMC \citep{Pietrzynski2019}. From the Galactic distributions, it is evident that there are both spectroscopically and photometrically identified C-rich Mira variables within the Galactic disc \emph{and} the Galactic bar-bulge. We also see a clump of stars associated with the Sgr dwarf spheroidal galaxy. In both samples, it appears there is a truncation in the radial distribution inside a radius of $\sim 5\,\mathrm{kpc}$ which may correspond to the corotation of the bar. Inside this radius, we observe an approximate barred structure aligned at approximately $20\,\mathrm{deg}$ with respect to the line-of-sight \citep[in agreement with other observations, e.g.][]{WeggGerhard2013,Simion2017}.

We isolate stars consistent in projection with bar-bulge membership as $|\ell|<20\,\mathrm{deg}$ and $|b|<20\,\mathrm{deg}$. These are shown as black points in the top right panel of Fig.~\ref{fig:crich}. We see three peaks in $W_{Ks,J-Ks}$ -- a foreground disc population, a bulge population and the Sgr stars. The separation between the bulge and Sgr peak is not particularly clean possibly due to background disc stars. Interestingly, the mean period is smallest in Sgr and largest in the foreground population. We will discuss this further below. We also display the C-rich Mira variables stars from \cite{Matsunaga2017} using their infrared photometry and the symbiotic star from \cite{Miszalski2013}. We isolate the Galactic bulge population as $\Delta s=|s-s_0|<3\,\mathrm{kpc}$ where $s_0$ is the distance to the Galactic centre. This region is shown in the left panels of Fig.~\ref{fig:crich}. There are $56$ spectroscopically and $269$ photometrically-identified C-rich stars in this region. In the lower right panel of Fig.~\ref{fig:crich} we show the median spectrum of these stars -- it is evident that they are predominantly C-rich.

In the lower row of Fig.~\ref{fig:crich}, the velocities of the C-rich stars in the on-sky bar-bulge region are shown where the members of Sgr are visible. We also display the transverse velocity distributions of the two samples. Fitting a Gaussian to the photometric sample we find dispersions of $132.9$ and $95.1\,\mathrm{km\,s}^{-1}$ in the longitudinal and latitudinal directions respectively (the proper motion uncertainties are of order $10\,\mathrm{km\,s}^{-1}$ so unimportant here). These dispersions are very similar to what is observed for the red clump giant stars \citep{Sanders2019a}. Also, the distributions have limited evidence of substructure. This then suggests that the C-rich stars are drawn from approximately the same population as the red clump giant stars and more generally the bulk bar-bulge population.

\subsection{C-rich bar-bulge member scenarios}
\begin{figure}
    \centering
    \includegraphics[width=\columnwidth]{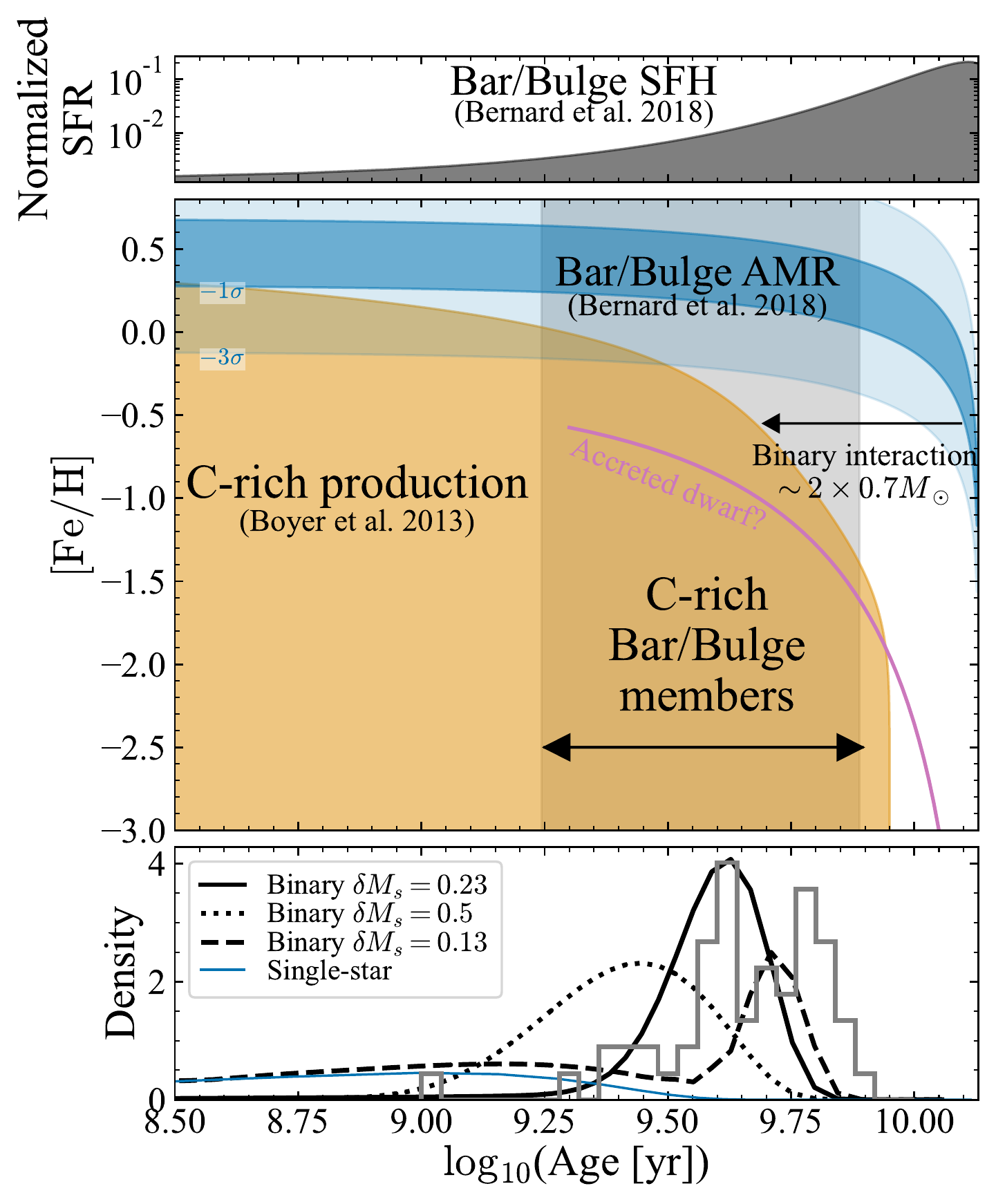}
    \caption{Likely scenarios for C-rich formation in the bar-bulge. The grey band shows the approximate age range of the spectroscopic C-rich candidates with the full distribution shown in the bottom panel. The orange-shaded region shows the range of age--metallicity combinations that give rise to C-rich TP-AGB production \protect\citep{Boyer2013}. The blue band shows the $\pm1$ and $3\sigma$ of the age--metallicity relation for the Galactic bar-bulge corresponding to the bar-bulge star formation history shown in the top panel \protect\citep{Bernard2018}. The pink line is an example age--metallicity relation for a dwarf galaxy that may have hypothetically merged into the bar-bulge. The horizontal leftwards arrow shows the approximate shift in remaining stellar age for the merger of two $0.7\,M_\odot$ stars -- in this way, binary interaction can produce old C-rich stars. The bottom panel shows the predictions from the single-star scenario in blue, and three binary scenario variants in black labelled by the width of the blue straggler mass distribution, $\delta M_s$.}
    \label{fig:scenarios}
\end{figure}
In the introduction we highlighted three possible reasons for C-rich bar-bulge stars: (i) there is recent (metal-poor) star formation in the Galactic bar-bulge, (i) they are accreted metal-poor stars, or (iii) the C-rich stars are formed primarily through binary interaction. These three scenarios are shown schematically in Fig.~\ref{fig:scenarios}. We discuss the evidence for each scenario in turn.

\subsubsection{In-situ bar-bulge star formation}\label{section::scenario_insitu}

The formation of C-rich stars through dredge-up is easier at lower metallicities as less carbon is required to counteract the already present oxygen. Lower mass stars have weaker dredge-up episodes meaning C-rich star production is a function of both mass and metallicity. Figure 8 from \cite{Boyer2013} shows that the upper limit in age at a given metallicity for the formation of C-rich stars is given approximately by
\begin{equation}
\log_{10}\frac{\tau_\mathrm{C}}{\mathrm{Gyr}}\approx0.95-\exp(1.3(\mathrm{[Fe/H]}-0.35)+0.8(\mathrm{[Fe/H]}+0.6)^3),
\end{equation}  as depicted in Fig.~\ref{fig:scenarios}. This means at metallicities of $\sim-2\,\mathrm{dex}$, C-rich stars can be as old as $\sim8\,\mathrm{Gyr}$ suggesting that C-rich bar-bulge stars could be remnants from the very earliest metal-poor phase of the bar-bulge region. However, the oldest C-rich Mira variables will also have the shortest periods. From C-rich Mira variables in the solar neighbourhood, \cite{Feast2006} concludes stars with $\log_{10}P\approx2.62$ have ages of $\sim2.5\,\mathrm{Gyr}$ which would correspond to masses of $\sim1.6M_\odot$. A compilation of literature results \citep{WyattCahn1983,FeastWhitelock1987,Eggen1998,FeastWhitelock2000,Feast2006,FeastWhitelock2014,Catchpole2016,LopezCorredoira2017,Grady2020,Nikzat2022,Sanders2022} suggests a simple approximation for the Mira variable period--age relation of $\tau\approx6.5(1+\tanh((330-P/\mathrm{d})/250))$ although recent theoretical relations \citep{Trabucchi2022} predict younger ages at fixed period. It is likely there is some metallicity dependence to the period--age relation for the Mira variables \citep{Trabucchi2022} but this is unlikely to make the ages at fixed periods significantly older than this. Utilising this relation, our sample of C-rich stars with $2.45<\log_{10}P/\mathrm{d}<2.75$ have ages between $1.7$ and $7.7\,\mathrm{Gyr}$ as shown by the band and the full distribution in Fig.~\ref{fig:scenarios}.

Early investigations of the bar-bulge star formation history concluded that it was predominantly an early $\sim10\,\mathrm{Gyr}$ old burst \citep{Zoccali2003}. However, evidence from the spectroscopic study of main-sequence turn-off stars \citep{Bensby2013} has pointed towards a small fraction of younger ($\sim3\,\mathrm{Gyr}$) stars. This younger minority population is supported by proper-motion-cleaned colour-magnitude diagrams \citep{Bernard2018} and corroborated by further age indicators \citep[as discussed by][]{Nataf2016}. In Fig.~\ref{fig:scenarios} we show the age-metallicity relation derived by \cite{Bernard2018} along with the star formation history they infer. We see that although there is a weak tail of star formation in the bar-bulge that extends to more recent times, the bar-bulge has enriched to on average super-solar metallicities by this time. The typical metallicity dispersion at each age is not well characterised, but from the results of \cite{Bernard2018} $0.2\,\mathrm{dex}$ is a reasonable value. We can estimate the fraction of C-rich stars within the Mira variable star population using
\begin{equation}
    \frac{N_\mathrm{C}}{N_\mathrm{total}} = \int_{t_\mathrm{min}}^{t_\mathrm{max}}\int_{-\infty}^\infty\mathrm{d}Z\mathrm{d}t\,\Gamma(t)\mathcal{N}(Z|Z_b(t),0.2)\Theta(\tau_\mathrm{C}(Z)-t),
\end{equation}
where $\Gamma(t)$ is the star-formation history and $\mathcal{N}(Z|Z_b(t),0.2)$ is a Gaussian with mean $Z_b(t)$ tracing the age-metallicity relation of the bar-bulge and dispersion $0.2\,\mathrm{dex}$ (both shown in Fig.~\ref{fig:scenarios}). ${t_\mathrm{min}}=1.07\,\mathrm{Gyr}$ and ${t_\mathrm{max}}=8.48\,\mathrm{Gyr}$ are the minimum and maximum ages corresponding to the observed period spread ($2.4<\log_{10}P<2.8$). $\Theta(x)$ is a Heaviside step function (evaluating to $1$ for $x>0$ and $0$ otherwise). This equation assumes that all stars are Mira variables for a similar time irrespective of the period. Although the TP-AGB phase is shorter for more massive stars, a higher fraction of this time is spent pulsating in the fundamental mode, so the relationship between Mira variable lifetime and mass is not simple \citep{Trabucchi2019}. Furthermore, we are assuming the two fields inspected by \cite{Bernard2018} are representative of the entire bar-bulge region. Using this relation, we find $N_\mathrm{C}/N_\mathrm{total}\approx1\times10^{-4}$. In the spectroscopically-classified sample, we have $N_\mathrm{C}/N_\mathrm{total}\approx3\times10^{-3}$. This theoretical calculation is slightly sensitive to the poorly-constrained low star-formation rate tail. Reasonable variations consistent with the star formation history from \citealt{Bernard2018} typically produce a factor of two variation in $N_\mathrm{C}/N_\mathrm{total}$ and to produce $N_\mathrm{C}/N_\mathrm{total}\approx3\times10^{-3}$ requires star formation histories strongly inconsistent with \cite{Bernard2018}. This then suggests that the star-formation history of the bar-bulge cannot explain the C-rich stars we observe. Furthermore, in the single-star model (blue line) the C-rich stars are predominantly skewed to younger ages/longer periods such that $N_\mathrm{C}/N_\mathrm{total}\approx5\times10^{-9}$ for $2.4<\log_{10}P/\mathrm{d}<2.6$ and the model would also predict significant numbers of C-rich stars with $\log_{10}P/\mathrm{d}>2.8$. This means the full age distribution of the C-rich Mira variables is a poor match to the data (see lower panel of Fig.~\ref{fig:scenarios}). However, this calculation also shows that the lack of C-rich stars for $\log_{10}P>2.75$ is putting strong limits on the star formation in the bar-bulge in the last $\sim\mathrm{Gyr}$, whilst from Fig.~\ref{fig:crich} we see the local disc stars have younger C-rich members.

The spatial and kinematic distributions suggest the C-rich population follows the bulk population in the bar-bulge despite being younger in the in situ formation scenario. However, the sample of microlensed dwarfs from \cite{Bensby2017} shows a similar extended distribution even for the young to intermediate-age stars. \cite{Debattista2017} argue that the spatial (and kinematic) distributions of different age/metallicity populations are a reflection of the different velocity dispersions of the populations prior to bar formation meaning different age populations should be distinguishable spatially and kinematically but younger populations still are anticipated to be present at higher latitude. This line of evidence alone does not completely rule against the in situ scenario. However, when combined with the predictions from the star formation history, the in situ star formation scenario is difficult to reconcile.

\subsubsection{Accreted metal-poor population}
We have observed that the bar-bulge population reaches too high metallicity at too early an epoch to explain the C-rich population observed. The next natural explanation is to invoke a more metal-poor star formation environment which subsequently merged into the bar-bulge region. For example, we have drawn a purely hypothetical age-metallicity track for a dwarf galaxy on Fig.~\ref{fig:scenarios}. A very ancient merger such as the suggested Kraken/Heracles \citep{Kruijssen,Horta} would lack C-rich stars, and the dwarf must have been accreted in the last $\sim5\,\mathrm{Gyr}$. The Sgr dwarf galaxy is a likely candidate here. The lower right panel of Fig.~\ref{fig:crich} shows all stars in the bar-bulge region of the sky coloured by proper motion. The Sgr dwarf is visible in proper motions with the suggestion there are other C-rich stars along the stream north of the Galactic plane. However, these stars are all at much further distances. The C-rich bar-bulge members are morphologically not similar to the Sgr distribution, nor do their kinematics suggest any association with Sgr. Another merger event that is perhaps more radial and more phase-mixed than Sgr is required. Inspecting Fig.~\ref{fig:crich} we see that on average the C-rich bar-bulge stars are longer period, or younger, than their counterparts in both Sgr and the LMC (the bar-bulge population has a mean of $\log_{10}P=2.6$, whilst the LMC and Sgr have $2.51$ and $2.55$ respectively). This then requires us to invoke a slightly peculiar star formation history for this suggested dwarf galaxy where there is only significant star formation recently. The problem is exacerbated if the dwarf galaxy is more metal-poor than Sgr and LMC. Furthermore, the minimum period of the bar-bulge C-rich Mira variables is also longer than that of the Sgr and LMC Mira variables. If we assume each group represents an approximately mono-metallicity population, Fig.~\ref{fig:scenarios} shows that a longer period minimum suggests a more metal-rich population. This suggests the progenitors of the bar-bulge C-rich Mira variables are more metal-rich than Sgr and LMC giving further evidence against an accreted population explanation. As discussed when considering Sgr, the spatial distribution (both on-sky and in 3D) and the kinematic distributions do not give any suggestion of being distinct from the broader bar-bulge population. Therefore, whilst there likely exist some merger configurations and star formation histories that could reproduce all observational constraints, the merger scenario explaining a significant number of the observed bar-bulge C-rich Mira variables does seem improbable.

\subsubsection{Binary channels}
In addition to the single star channels, C-rich stars can form through a binary channel. Binary mass transfer increases the mass of the secondary potentially to the extent that it is of high enough mass to later become a C-rich star. In extreme cases, a stellar collision can approximately double the mass of a star. If the primary companion is itself a C-rich star, it may require lower mass transfer to make the secondary C-rich. In Fig.~\ref{fig:scenarios} we show the shift in apparent age produced by the merger of two $\sim12\,\mathrm{Gyr}$ old $0.7M_\odot$ stars (assuming $\tau\propto M^{-2.5}$). At metallicity $\sim-0.5\,\mathrm{dex}$ this change in mass is sufficient to bring the star into the C-rich formation region. These binary products would first appear as blue straggler stars before eventually evolving to C-rich stars through dredge-up.

The production of C-rich Mira variables in old environments is evidenced by the presence of a C-rich Mira variable in the globular cluster Lyng\aa\ 7 \citep{FeastLynga}. Its radial velocity is consistent with membership of the globular cluster although the Gaia DR3 proper motion measurement is inconsistent possibly due to contamination in the cluster environment (there are two nearby sources with only 2-parameter astrometric solutions). \cite{FeastLynga} hypothesised that this star was formed through the collision/merger of two $\sim0.8M_\odot$ stars producing a blue straggler star which subsequently evolved to be a C-rich Mira variable. There is the suggestion that $\sim27\percent$ of bar-bulge stars were formed in globular clusters \citep{Horta2021} such that it is possible any blue stragglers in the bar-bulge are in fact the result of cluster evolution. However, the typical timescale for blue stragglers to survive is $\sim1\,\mathrm{Gyr}$ meaning we will only be sensitive to cluster evolution products that formed in a cluster that very recently dissolved. None of the identified stars appears to be associated with globular clusters (the minimum separation relative to the \citealt{Harris2010} globular cluster list is $0.35\,\mathrm{deg}$ for the spectroscopic classifications and $0.17\,\mathrm{deg}$ for the photometric classifications). Globular clusters show no correlation between blue straggler fraction and density \citep{Knigge} suggesting binary evolution rather than collisions form the majority of blue stragglers in older systems. This is evidenced by the presence of blue stragglers in the field \citep[e.g][]{Carney2001}. Complete mergers of close binary systems, rather than collisions in dense environments, are also a subdominant channel with old clusters producing $\lesssim20\percent$ of blue stragglers via this channel \citep{Geller2013,LeinerGeller}. This suggests mass transfer is the dominant blue straggler production channel in old clusters and in particular in the field.

Although the binary fraction is lower in denser environments \citep{Milone2012}, the products of binary evolution have been observed in the bar-bulge.
\cite{Clarkson2011} discovered $\sim30$ blue straggler bar-bulge members using proper-motion-cleaned colour-magnitude diagrams and photometric lightcurves in the Hubble Space Telescope SWEEPS field. They optimistically classify $29-37$ stars as blue stragglers and more conservatively $18-22$ depending on the assumption of a young bar-bulge population. There is also evidence of carbon-enhanced metal-poor stars and CH stars in the bar-bulge, although potentially at a lower fraction than the local disc fraction possibly due to the binary fraction variation with metallicity or density \citep{ArentsenPIGS}. \cite{Azzopardi1988, Azzopardi1991} discovered a series of C-rich giant stars towards the bar-bulge that are too faint to be AGB stars so are likely products of binary evolution that could go on to be C-rich Mira variables.

Recently, \cite{Marigo2022} has studied the occurrence of C-rich TP-AGB in open clusters using the more reliable membership probabilities now possible using Gaia. They concluded that for the intermediate age ($\sim1.5\,\mathrm{Gyr}$) clusters NGC 7789 and NGC 2660, the single star channel produces $\sim10-1000$ more C-rich TP-AGB stars than the binary channel (by anchoring to the observed number of blue stragglers in each cluster). Following the calculation in \cite{Marigo2022}, we can relate the observed number of blue stragglers in the bar-bulge to the expected number of C-rich Mira variables as
\begin{equation}
\frac{N_\mathrm{C}}{N_\mathrm{BSS}} = \frac{\int\mathrm{d}M\mathrm{d}t\mathrm{d}Z\,\Gamma(t)p(Z)\tau_\mathrm{Mira}(M) p(M|t)\Theta(\tau_\mathrm{C}(Z)-\tau(M))}{\int_{1.41M_\odot}^{2.11M_\odot}\mathrm{d}M\mathrm{d}t\,\Gamma(t)\tau_\mathrm{MS}(M) p(M|t)}
\label{eqn:nc_nbss}
\end{equation}
where $p(Z)=\mathcal{N}(Z|Z_b(t),0.2)$. The integration ranges in the numerator cover all valid $Z$ and $t$ (up to $\sim13\,\mathrm{Gyr}$) and from the main sequence turn-off mass up to infinity for $M$. In the denominator, we again consider all valid $t$ but restrict to only considering blue-stragglers with masses $1.41<M/M_\odot<2.11$ as \cite{Clarkson2011} reports only being sensitive to these blue stragglers. We assume the relationship between main sequence age and mass is simply $\tau(M)\approx10\,\mathrm{Gyr}(M/M_\odot)^{-2.5}$. $\tau_\mathrm{Mira}(M)$ is the approximate lifetime of the Mira phase which we assume is a constant $0.2\,\mathrm{Myr}$ based on the results from \cite{Trabucchi2019}. $p(M|t)$ is the probability of producing a blue straggler of mass $M$ in a population of age $t$. Both \cite{LeinerGeller} and \cite{Jadhav} provide estimates for this distribution in terms of the mass in excess of the main sequence turn-off mass, $\delta M$, based on results from Gaia for clusters. We fit an approximate half-Gaussian centred on zero to these distributions (the plotted $7\,\mathrm{Gyr}$ distribution from \citealt{LeinerGeller} and table 1 from \citealt{Jadhav} for the $9.75-10$ log(age) clusters) finding a standard deviation of $\delta M_s\approx0.5M_\odot$. Assuming a constant remaining blue straggler lifetime $\tau_\mathrm{MS}(M)$ with mass $M$, we find this calculation yields $N_\mathrm{C}/N_\mathrm{BSS}=5.4\times10^{-5}/(\tau_\mathrm{MS}/\,\mathrm{Gyr})$. The largest uncertainty arises from the remaining blue straggler lifetime. \cite{LeinerGeller} consider several models for binary mass transfer finding the L2/L3 overflow model produces the best match to the cluster blue straggler distribution although not completely reproducing all features. For an old $7\,\mathrm{Gyr}$ population, the remaining main sequence lifetime from this model ranges from $400\,\mathrm{Myr}$ to $4\,\mathrm{Gyr}$ depending on the mass ratio. We adopt $\tau_\mathrm{MS}\approx1\,\mathrm{Gyr}$ giving $N_\mathrm{C}/N_\mathrm{BSS}=5.4\times10^{-5}$ but note a factor $\sim2$ uncertainty in this number.

To compare with the number of blue stragglers found by \citet[][$N_\mathrm{BSS,C11}$]{Clarkson2011} we normalize by the respective stellar masses contained in the two areas considered:
\begin{equation}
    \frac{A_\mathrm{SWEEPS}}{A_\mathrm{Bar-Bulge}} = \frac{\int_{\ell_\mathrm{SWEEPS}-\Delta\ell}^{\ell_\mathrm{SWEEPS}+\Delta\ell}\mathrm{d}\ell\,\int_{b_\mathrm{SWEEPS}-\Delta b}^{b_\mathrm{SWEEPS}+\Delta b}\mathrm{d}b\,\mathrm{d}\ell\cos b\rho(\ell,b)}{\int^{\ell_\mathrm{max}}_{-\ell_\mathrm{max}}\mathrm{d}\ell\,\int^{\pi/2}_{b_\mathrm{min}}\mathrm{d}b\,\mathrm{d}\ell\cos b\rho(\ell,b)}
\end{equation}
where $\rho(\ell,b)$ is the bar-bulge density profile approximated as an exponential in $\ell$ and $b$ with scalelengths of $3.5\,\mathrm{deg}$ and $1.3\,\mathrm{deg}$ respectively \citep{WeggGerhard2013}. The density profile is integrated from $-\ell_\mathrm{max}=-20$ to $\ell_\mathrm{max}=20\,\mathrm{deg}$ in $\ell$ and for $b>b_\mathrm{min}=1\,\mathrm{deg}$ as extinction reduces the density of BP/RP C-rich detections from Gaia below this latitude. This calculation gives $A_\mathrm{SWEEPS}/A_\mathrm{Bar-Bulge}\approx3.5\times10^{-5}$. Expanding to the full bar-bulge gives $A_\mathrm{SWEEPS}/A_\mathrm{Bar-Bulge}\approx1.5\times10^{-5}$ i.e. $2.3$ times more C-rich stars whereas we find $\sim5$ times more stars. This might reflect more contamination in the photometric samples or an inappropriate density law employed for the low-latitude regions. We estimate the expected number of bar-bulge spectroscopic C-rich Mira variables as $N_\mathrm{C}=N_\mathrm{BSS, C11}(N_\mathrm{C}/N_\mathrm{BSS})/(A_\mathrm{SWEEPS}/A_\mathrm{Bar-Bulge})\approx44$. This very nicely matches the $56$ observed spectroscopic C-rich Mira variables but as discussed the uncertainty on the estimate is probably around a factor $2-3$ as we can vary the blue straggler mass distribution (as discussed below), the remaining main sequence lifetime of blue stragglers, the lifetime of Mira variable stars and the specifics of the density modelling.

The advantage of this channel relative to the in-situ formation channel described in Section~\ref{section::scenario_insitu} is that the peak of the predicted period distribution shifts to lower periods. We compute the expected period distribution by not integrating over $M$ in the numerator of equation~\eqref{eqn:nc_nbss} to find the blue straggler mass distribution. We convert this mass distribution into effective age ($\tau\approx(10\,\mathrm{Gyr})(M/M_\odot)^{-2.5}$) and period distributions as shown in Fig.~\ref{fig:scenarios} using the relation given in Section~\ref{section::scenario_insitu} and the appropriate Jacobians. The mode of the distribution is $\log_{10}P\approx2.70$ with width $0.05\,\mathrm{dex}$. As shown in the lower panel of Fig.~\ref{fig:scenarios}, this is not a particularly good match to the data which has median $\log_{10}P$ of $\sim2.61$ and width $0.07$. The location of the peak period is a balance of increased mass to produce more C-rich stars at fixed metallicity whilst keeping the mass low enough to not overly bias towards longer period (younger) stars. It is the lower mass blue-stragglers that contribute to the lowering of the period distribution so to produce a lower mean period we must narrow the blue straggler mass distribution width, a not unreasonable suggestion considering the uncertainties and the use of a perhaps inappropriate $p(M)$ based on cluster stars. When we narrow the standard deviation of $p(M)$ to $\delta M_s=0.13M_\odot$ as shown in Fig.~\ref{fig:scenarios}, we produce a high effective age (shorter period) peak from the blue straggler stars with masses greater than the turn-off mass, along with a broader low effective age (longer period) peak from stars with masses around the turn-off mass that approximately resembles the single-star distribution discussed in Section~\ref{section::scenario_insitu}. Seeking a compromise we set the standard deviation of the $p(M)$ distribution as $\delta M_s=0.23M_\odot$ (solid black line in Fig.~\ref{fig:scenarios}) and find a better match to the data with the mode of the distribution at $\log_{10}P\approx2.64$ with width $0.09\,\mathrm{dex}$. This choice reduces the expected number of C-rich Mira variables to around $20$ but again the other uncertainties are large.

We have demonstrated that the binary channel can reproduce the observed number of C-rich Mira variables under reasonable assumptions and that it provides a better match to the period distribution than the single-star channel. Furthermore, of the three considered scenarios, the binary channel scenario is perhaps most consistent with the observation that both the spatial and kinematic distributions of the C-rich stars are very similar to that of the red clump giant stars in the Galactic bar-bulge. The red clump star distribution predominantly traces the properties of the bar-bulge for stars formed around the peak of star formation with a small bias towards red clump stars preferentially being found in younger populations. All these lines of evidence then indicate that the bulk of our observed bar-bulge C-rich Mira variable sample is likely formed through binary evolution.

\section{Conclusions}
The separation of O-rich and C-rich long-period variables is crucial for their precision use as distance tracers and indicators of the age/metallicity of stellar populations. Here we have demonstrated the power of the Gaia BP/RP spectra for this task. Using a simple unsupervised approach based on the Uniform Manifold Approximation and Projection (UMAP) algorithm, we have naturally identified two broad groups of spectra that are associated with O-rich and C-rich objects. We have discussed how the unsupervised approach can be used to learn about the nature of the stars beyond their simple O/C separation, in particular how we can find some S-stars and also emission line objects that are possibly symbiotic. We have demonstrated how utilising the information from the entire spectrum offers an improvement in the classification over simpler diagnostics. Our classification scheme has been further validated on the basis of infrared colour-colour diagrams and we have shown that a supervised scheme using Gaia and 2MASS photometry and the unsupervised classifications offers an improvement over simpler colour-colour cuts.

Using both the spectroscopic/unsupervised and photometric/supervised classifications we have identified a small population of C-rich stars in the Galactic bar-bulge region. Their spatial and kinematic distributions are in agreement with other bar-bulge tracers such as red clump giants suggesting they are an in-situ population associated with the bulk of the bar-bulge. Their production via single-star evolution typically produces a factor of ten too few stars than observed because of the bar-bulge's predominantly early episodes of star formation that quickly enriches the inner Galaxy to high metallicity. Old high metallicity populations do not form C-rich stars. If we instead consider these stars as the products of binary evolution, we expect them to be the evolved versions of blue straggler stars. A rather simple model of the blue straggler production in the bar-bulge approximately reproduces the period distribution of our sample and the observed number of C-rich Mira variables across the entire bar-bulge when referencing against the observed number of blue-straggler stars in the SWEEPS field. This demonstrates that the entire population of C-rich Mira variables can be attributed to binary evolution and there is limited evidence for a significant young in-situ or accreted population.

Note that we have restricted our analysis to the long-period variables as the Mira variables are of particular interest for Galactic and cosmological studies. However, our analysis is simply extended to all stars in Gaia. Indeed it would be interesting to perform a dimensionality-reduction analysis on the entire BP/RP dataset to identify and separate the gross stellar types but to also identify unusual outlier groups of stars, galaxies or quasars.

\section*{Data Availability}
All data used in this work are in the public domain. We have made our classifications and the UMAP coordinates for the full dataset available here: \url{https://www.homepages.ucl.ac.uk/~ucapjls/data/gaia_dr3_lpv_classifications.fits}.

\section*{Acknowledgements}
{
JLS thanks the support of the Royal Society (URF\textbackslash R1\textbackslash191555), the hospitality of the Flatiron Institute and the organisers of the Gaia F\^ete where this project was started. We thank Jo Ciuc\u{a} for useful conversations in the preparation of this work. This paper made use of the Whole Sky Database (wsdb) created by Sergey Koposov and maintained at the Institute of Astronomy, Cambridge by Sergey Koposov, Vasily Belokurov and Wyn Evans with financial support from the Science \& Technology Facilities Council (STFC) and the European Research Council (ERC). This software made use of the Q3C software \citep{q3c}. This research has made use of the SVO Filter Profile Service (\url{http://svo2.cab.inta-csic.es/theory/fps/}) supported from the Spanish MINECO through grant AYA2017-84089. This research has made use of the SIMBAD database, operated at CDS, Strasbourg, France.
This work has made use of data from the European Space Agency (ESA) mission
{\it Gaia} (\url{https://www.cosmos.esa.int/gaia}), processed by the {\it Gaia}
Data Processing and Analysis Consortium (DPAC,
\url{https://www.cosmos.esa.int/web/gaia/dpac/consortium}). Funding for the DPAC
has been provided by national institutions, in particular the institutions
participating in the {\it Gaia} Multilateral Agreement.
This publication makes use of data products from the Two Micron All Sky Survey, which is a joint project of the University of Massachusetts and the Infrared Processing and Analysis Center/California Institute of Technology, funded by the National Aeronautics and Space Administration and the National Science Foundation.
This research has made use of the International Variable Star Index (VSX) database, operated at AAVSO, Cambridge, Massachusetts, USA. This paper made use of
\textsc{numpy} \citep{numpy},
\textsc{scipy} \citep{scipy},
\textsc{matplotlib} \citep{matplotlib},
\textsc{seaborn} \citep{seaborn}, \textsc{pandas} \citep{pandas1}
\textsc{astropy} \citep{astropy:2013,astropy:2018} and \textsc{vaex} \citep{vaex}.
}


\bibliographystyle{mnras}
\bibliography{bibliography}

\begin{thebibliography}{}
\makeatletter
\relax
\def\mn@urlcharsother{\let\do\@makeother \do\$\do\&\do\#\do\^\do\_\do\%\do\~}
\def\mn@doi{\begingroup\mn@urlcharsother \@ifnextchar [ {\mn@doi@}
  {\mn@doi@[]}}
\def\mn@doi@[#1]#2{\def\@tempa{#1}\ifx\@tempa\@empty \href
  {http://dx.doi.org/#2} {doi:#2}\else \href {http://dx.doi.org/#2} {#1}\fi
  \endgroup}
\def\mn@eprint#1#2{\mn@eprint@#1:#2::\@nil}
\def\mn@eprint@arXiv#1{\href {http://arxiv.org/abs/#1} {{\tt arXiv:#1}}}
\def\mn@eprint@dblp#1{\href {http://dblp.uni-trier.de/rec/bibtex/#1.xml}
  {dblp:#1}}
\def\mn@eprint@#1:#2:#3:#4\@nil{\def\@tempa {#1}\def\@tempb {#2}\def\@tempc
  {#3}\ifx \@tempc \@empty \let \@tempc \@tempb \let \@tempb \@tempa \fi \ifx
  \@tempb \@empty \def\@tempb {arXiv}\fi \@ifundefined
  {mn@eprint@\@tempb}{\@tempb:\@tempc}{\expandafter \expandafter \csname
  mn@eprint@\@tempb\endcsname \expandafter{\@tempc}}}

\bibitem[\protect\citeauthoryear{{Aaronson}, {Blanco}, {Cook}, {Olszewski}  \&
  {Schechter}}{{Aaronson} et~al.}{1990}]{Aaronson1990}
{Aaronson} M.,  {Blanco} V.~M.,  {Cook} K.~H.,  {Olszewski} E.~W.,
  {Schechter} P.~L.,  1990, \mn@doi [\apjs] {10.1086/191491}, \href
  {https://ui.adsabs.harvard.edu/abs/1990ApJS...73..841A} {73, 841}

\bibitem[\protect\citeauthoryear{{Alonso-Garc{\'\i}a}
  et~al.,}{{Alonso-Garc{\'\i}a} et~al.}{2017}]{AlonsoGarcia2017}
{Alonso-Garc{\'\i}a} J.,  et~al., 2017, \mn@doi [\apjl]
  {10.3847/2041-8213/aa92c3}, \href
  {https://ui.adsabs.harvard.edu/abs/2017ApJ...849L..13A} {849, L13}

\bibitem[\protect\citeauthoryear{{Anders}, {Chiappini}, {Santiago},
  {Matijevi{\v{c}}}, {Queiroz}, {Steinmetz}  \& {Guiglion}}{{Anders}
  et~al.}{2018}]{Anders2018}
{Anders} F.,  {Chiappini} C.,  {Santiago} B.~X.,  {Matijevi{\v{c}}} G.,
  {Queiroz} A.~B.,  {Steinmetz} M.,   {Guiglion} G.,  2018, \mn@doi [\aap]
  {10.1051/0004-6361/201833099}, \href
  {https://ui.adsabs.harvard.edu/abs/2018A&A...619A.125A} {619, A125}

\bibitem[\protect\citeauthoryear{{Andrae} et~al.,}{{Andrae}
  et~al.}{2022}]{Andrae2022}
{Andrae} R.,  et~al., 2022, \mn@doi [arXiv e-prints]
  {10.48550/arXiv.2206.06138}, \href
  {https://ui.adsabs.harvard.edu/abs/2022arXiv220606138A} {p. arXiv:2206.06138}

\bibitem[\protect\citeauthoryear{{Arentsen} et~al.,}{{Arentsen}
  et~al.}{2021}]{ArentsenPIGS}
{Arentsen} A.,  et~al., 2021, \mn@doi [\mnras] {10.1093/mnras/stab1343}, \href
  {https://ui.adsabs.harvard.edu/abs/2021MNRAS.505.1239A} {505, 1239}

\bibitem[\protect\citeauthoryear{{Astropy Collaboration} et~al.,}{{Astropy
  Collaboration} et~al.}{2013}]{astropy:2013}
{Astropy Collaboration} et~al., 2013, \mn@doi [\aap]
  {10.1051/0004-6361/201322068}, \href
  {http://adsabs.harvard.edu/abs/2013A%26A...558A..33A} {558, A33}

\bibitem[\protect\citeauthoryear{{Azzopardi}, {Lequeux}  \&
  {Rebeirot}}{{Azzopardi} et~al.}{1988}]{Azzopardi1988}
{Azzopardi} M.,  {Lequeux} J.,   {Rebeirot} E.,  1988, \aap, \href
  {https://ui.adsabs.harvard.edu/abs/1988A&A...202L..27A} {202, L27}

\bibitem[\protect\citeauthoryear{{Azzopardi}, {Lequeux}, {Rebeirot}  \&
  {Westerlund}}{{Azzopardi} et~al.}{1991}]{Azzopardi1991}
{Azzopardi} M.,  {Lequeux} J.,  {Rebeirot} E.,   {Westerlund} B.~E.,  1991,
  \aaps, \href {https://ui.adsabs.harvard.edu/abs/1991A&AS...88..265A} {88,
  265}

\bibitem[\protect\citeauthoryear{{Belokurov}, {Erkal}, {Deason}, {Koposov}, {De
  Angeli}, {Evans}, {Fraternali}  \& {Mackey}}{{Belokurov}
  et~al.}{2017}]{Belokurov2016}
{Belokurov} V.,  {Erkal} D.,  {Deason} A.~J.,  {Koposov} S.~E.,  {De Angeli}
  F.,  {Evans} D.~W.,  {Fraternali} F.,   {Mackey} D.,  2017, \mn@doi [\mnras]
  {10.1093/mnras/stw3357}, \href
  {https://ui.adsabs.harvard.edu/abs/2017MNRAS.466.4711B} {466, 4711}

\bibitem[\protect\citeauthoryear{{Belokurov}, {Vasiliev}, {Deason}, {Koposov},
  {Fattahi}, {Dillamore}, {Davies}  \& {Grand}}{{Belokurov}
  et~al.}{2022}]{Belokurov2022}
{Belokurov} V.,  {Vasiliev} E.,  {Deason} A.~J.,  {Koposov} S.~E.,  {Fattahi}
  A.,  {Dillamore} A.~M.,  {Davies} E.~Y.,   {Grand} R. J.~J.,  2022, arXiv
  e-prints, \href {https://ui.adsabs.harvard.edu/abs/2022arXiv220811135B} {p.
  arXiv:2208.11135}

\bibitem[\protect\citeauthoryear{{Bensby} et~al.,}{{Bensby}
  et~al.}{2013}]{Bensby2013}
{Bensby} T.,  et~al., 2013, \mn@doi [\aap] {10.1051/0004-6361/201220678}, \href
  {https://ui.adsabs.harvard.edu/abs/2013A&A...549A.147B} {549, A147}

\bibitem[\protect\citeauthoryear{{Bensby} et~al.,}{{Bensby}
  et~al.}{2017}]{Bensby2017}
{Bensby} T.,  et~al., 2017, \mn@doi [\aap] {10.1051/0004-6361/201730560}, \href
  {https://ui.adsabs.harvard.edu/abs/2017A&A...605A..89B} {605, A89}

\bibitem[\protect\citeauthoryear{{Bernard}, {Schultheis}, {Di Matteo}, {Hill},
  {Haywood}  \& {Calamida}}{{Bernard} et~al.}{2018}]{Bernard2018}
{Bernard} E.~J.,  {Schultheis} M.,  {Di Matteo} P.,  {Hill} V.,  {Haywood} M.,
   {Calamida} A.,  2018, \mn@doi [\mnras] {10.1093/mnras/sty902}, \href
  {https://ui.adsabs.harvard.edu/abs/2018MNRAS.477.3507B} {477, 3507}

\bibitem[\protect\citeauthoryear{{Bhardwaj} et~al.,}{{Bhardwaj}
  et~al.}{2019}]{Bhardwaj2019}
{Bhardwaj} A.,  et~al., 2019, \mn@doi [\apj] {10.3847/1538-4357/ab38c2}, \href
  {https://ui.adsabs.harvard.edu/abs/2019ApJ...884...20B} {884, 20}

\bibitem[\protect\citeauthoryear{{Blanco}, {McCarthy}  \& {Blanco}}{{Blanco}
  et~al.}{1984}]{Blanco1984}
{Blanco} V.~M.,  {McCarthy} M.~F.,   {Blanco} B.~M.,  1984, \mn@doi [\aj]
  {10.1086/113560}, \href
  {https://ui.adsabs.harvard.edu/abs/1984AJ.....89..636B} {89, 636}

\bibitem[\protect\citeauthoryear{{Bobrovnikoff}}{{Bobrovnikoff}}{1933}]{Bobrovnikoff1933}
{Bobrovnikoff} N.~T.,  1933, \mn@doi [\apj] {10.1086/143501}, \href
  {https://ui.adsabs.harvard.edu/abs/1933ApJ....78..211B} {78, 211}

\bibitem[\protect\citeauthoryear{{Bovy}, {Leung}, {Hunt}, {Mackereth},
  {Garc{\'\i}a-Hern{\'a}ndez}  \& {Roman-Lopes}}{{Bovy}
  et~al.}{2019}]{Bovy2019}
{Bovy} J.,  {Leung} H.~W.,  {Hunt} J. A.~S.,  {Mackereth} J.~T.,
  {Garc{\'\i}a-Hern{\'a}ndez} D.~A.,   {Roman-Lopes} A.,  2019, \mn@doi
  [\mnras] {10.1093/mnras/stz2891}, \href
  {https://ui.adsabs.harvard.edu/abs/2019MNRAS.490.4740B} {490, 4740}

\bibitem[\protect\citeauthoryear{{Boyer} et~al.,}{{Boyer}
  et~al.}{2013}]{Boyer2013}
{Boyer} M.~L.,  et~al., 2013, \mn@doi [\apj] {10.1088/0004-637X/774/1/83},
  \href {https://ui.adsabs.harvard.edu/abs/2013ApJ...774...83B} {774, 83}

\bibitem[\protect\citeauthoryear{{Breddels} \& {Veljanoski}}{{Breddels} \&
  {Veljanoski}}{2018}]{vaex}
{Breddels} M.~A.,  {Veljanoski} J.,  2018, \mn@doi [\aap]
  {10.1051/0004-6361/201732493}, \href
  {https://ui.adsabs.harvard.edu/abs/2018A&A...618A..13B} {618, A13}

\bibitem[\protect\citeauthoryear{{Brewer}, {Richer}  \& {Crabtree}}{{Brewer}
  et~al.}{1995}]{Brewer1996}
{Brewer} J.~P.,  {Richer} H.~B.,   {Crabtree} D.~R.,  1995, \mn@doi [\aj]
  {10.1086/117466}, \href
  {https://ui.adsabs.harvard.edu/abs/1995AJ....109.2480B} {109, 2480}

\bibitem[\protect\citeauthoryear{{Carney}, {Latham}, {Laird}, {Grant}  \&
  {Morse}}{{Carney} et~al.}{2001}]{Carney2001}
{Carney} B.~W.,  {Latham} D.~W.,  {Laird} J.~B.,  {Grant} C.~E.,   {Morse}
  J.~A.,  2001, \mn@doi [\aj] {10.1086/324233}, \href
  {https://ui.adsabs.harvard.edu/abs/2001AJ....122.3419C} {122, 3419}

\bibitem[\protect\citeauthoryear{{Carrasco} et~al.,}{{Carrasco}
  et~al.}{2021}]{Carrasco2021}
{Carrasco} J.~M.,  et~al., 2021, \mn@doi [\aap] {10.1051/0004-6361/202141249},
  \href {https://ui.adsabs.harvard.edu/abs/2021A&A...652A..86C} {652, A86}

\bibitem[\protect\citeauthoryear{{Catchpole}, {Whitelock}, {Feast}, {Hughes},
  {Irwin}  \& {Alard}}{{Catchpole} et~al.}{2016}]{Catchpole2016}
{Catchpole} R.~M.,  {Whitelock} P.~A.,  {Feast} M.~W.,  {Hughes} S. M.~G.,
  {Irwin} M.,   {Alard} C.,  2016, \mn@doi [\mnras] {10.1093/mnras/stv2372},
  \href {https://ui.adsabs.harvard.edu/abs/2016MNRAS.455.2216C} {455, 2216}

\bibitem[\protect\citeauthoryear{Chen \& Guestrin}{Chen \&
  Guestrin}{2016}]{xgboost}
Chen T.,  Guestrin C.,  2016, in Proceedings of the 22nd ACM SIGKDD
  International Conference on Knowledge Discovery and Data Mining. KDD '16.
ACM, New York, NY, USA, pp 785--794, \mn@doi{10.1145/2939672.2939785}, \url
  {http://doi.acm.org/10.1145/2939672.2939785}

\bibitem[\protect\citeauthoryear{{Chen}, {Trager}, {Peletier}, {Lan{\c{c}}on},
  {Vazdekis}, {Prugniel}, {Silva}  \& {Gonneau}}{{Chen}
  et~al.}{2014}]{Chen2014}
{Chen} Y.-P.,  {Trager} S.~C.,  {Peletier} R.~F.,  {Lan{\c{c}}on} A.,
  {Vazdekis} A.,  {Prugniel} P.,  {Silva} D.~R.,   {Gonneau} A.,  2014, \mn@doi
  [\aap] {10.1051/0004-6361/201322505}, \href
  {https://ui.adsabs.harvard.edu/abs/2014A&A...565A.117C} {565, A117}

\bibitem[\protect\citeauthoryear{{Clarkson} et~al.,}{{Clarkson}
  et~al.}{2011}]{Clarkson2011}
{Clarkson} W.~I.,  et~al., 2011, \mn@doi [\apj] {10.1088/0004-637X/735/1/37},
  \href {https://ui.adsabs.harvard.edu/abs/2011ApJ...735...37C} {735, 37}

\bibitem[\protect\citeauthoryear{{Creevey} et~al.,}{{Creevey}
  et~al.}{2022}]{Creevey2022APSIS}
{Creevey} O.~L.,  et~al., 2022, \mn@doi [arXiv e-prints]
  {10.48550/arXiv.2206.05864}, \href
  {https://ui.adsabs.harvard.edu/abs/2022arXiv220605864C} {p. arXiv:2206.05864}

\bibitem[\protect\citeauthoryear{{De Angeli} et~al.,}{{De Angeli}
  et~al.}{2022}]{DeAngeli2022}
{De Angeli} F.,  et~al., 2022, arXiv e-prints, \href
  {https://ui.adsabs.harvard.edu/abs/2022arXiv220606143D} {p. arXiv:2206.06143}

\bibitem[\protect\citeauthoryear{{De Marco} \& {Izzard}}{{De Marco} \&
  {Izzard}}{2017}]{DeMarco2017}
{De Marco} O.,  {Izzard} R.~G.,  2017, \mn@doi [\pasa] {10.1017/pasa.2016.52},
  \href {https://ui.adsabs.harvard.edu/abs/2017PASA...34....1D} {34, e001}

\bibitem[\protect\citeauthoryear{{Deason}, {Belokurov}, {Erkal}, {Koposov}  \&
  {Mackey}}{{Deason} et~al.}{2017}]{Deason2017}
{Deason} A.~J.,  {Belokurov} V.,  {Erkal} D.,  {Koposov} S.~E.,   {Mackey} D.,
  2017, \mn@doi [\mnras] {10.1093/mnras/stx263}, \href
  {https://ui.adsabs.harvard.edu/abs/2017MNRAS.467.2636D} {467, 2636}

\bibitem[\protect\citeauthoryear{{Debattista}, {Ness}, {Gonzalez}, {Freeman},
  {Zoccali}  \& {Minniti}}{{Debattista} et~al.}{2017}]{Debattista2017}
{Debattista} V.~P.,  {Ness} M.,  {Gonzalez} O.~A.,  {Freeman} K.,  {Zoccali}
  M.,   {Minniti} D.,  2017, \mn@doi [\mnras] {10.1093/mnras/stx947}, \href
  {https://ui.adsabs.harvard.edu/abs/2017MNRAS.469.1587D} {469, 1587}

\bibitem[\protect\citeauthoryear{{Eggen}}{{Eggen}}{1998}]{Eggen1998}
{Eggen} O.~J.,  1998, \mn@doi [\aj] {10.1086/300354}, \href
  {https://ui.adsabs.harvard.edu/abs/1998AJ....115.2435E} {115, 2435}

\bibitem[\protect\citeauthoryear{Ester, Kriegel, Sander  \& Xu}{Ester
  et~al.}{1996}]{DBSCAN}
Ester M.,  Kriegel H.-P.,  Sander J.,   Xu X.,  1996, in Proceedings of the
  Second International Conference on Knowledge Discovery and Data Mining.
  KDD'96.
AAAI Press, p. 226–231

\bibitem[\protect\citeauthoryear{{Evans} et~al.,}{{Evans}
  et~al.}{2018}]{Evans2018}
{Evans} D.~W.,  et~al., 2018, \mn@doi [\aap] {10.1051/0004-6361/201832756},
  \href {https://ui.adsabs.harvard.edu/abs/2018A&A...616A...4E} {616, A4}

\bibitem[\protect\citeauthoryear{{Feast} \& {Whitelock}}{{Feast} \&
  {Whitelock}}{1987}]{FeastWhitelock1987}
{Feast} M.~W.,  {Whitelock} P.~A.,  1987, in {Kwok} S.,  {Pottasch} S.~R.,
  eds, Late Stages of Stellar Evolution. p.~33,
  \mn@doi{10.1007/978-94-009-3813-7\_3}

\bibitem[\protect\citeauthoryear{{Feast} \& {Whitelock}}{{Feast} \&
  {Whitelock}}{2000}]{FeastWhitelock2000}
{Feast} M.~W.,  {Whitelock} P.~A.,  2000, \mn@doi [\mnras]
  {10.1046/j.1365-8711.2000.03629.x}, \href
  {https://ui.adsabs.harvard.edu/abs/2000MNRAS.317..460F} {317, 460}

\bibitem[\protect\citeauthoryear{{Feast} \& {Whitelock}}{{Feast} \&
  {Whitelock}}{2014}]{FeastWhitelock2014}
{Feast} M.,  {Whitelock} P.~A.,  2014, in {Feltzing} S.,  {Zhao} G.,  {Walton}
  N.~A.,   {Whitelock} P.,  eds,  IAU Symposium Vol. 298, Setting the scene for
  Gaia and LAMOST. pp 40--52 (\mn@eprint {arXiv} {1310.3928}),
  \mn@doi{10.1017/S1743921313006182}

\bibitem[\protect\citeauthoryear{{Feast}, {Glass}, {Whitelock}  \&
  {Catchpole}}{{Feast} et~al.}{1989}]{Feast1989}
{Feast} M.~W.,  {Glass} I.~S.,  {Whitelock} P.~A.,   {Catchpole} R.~M.,  1989,
  \mn@doi [\mnras] {10.1093/mnras/241.3.375}, \href
  {https://ui.adsabs.harvard.edu/abs/1989MNRAS.241..375F} {241, 375}

\bibitem[\protect\citeauthoryear{{Feast}, {Whitelock}  \& {Menzies}}{{Feast}
  et~al.}{2006}]{Feast2006}
{Feast} M.~W.,  {Whitelock} P.~A.,   {Menzies} J.~W.,  2006, \mn@doi [\mnras]
  {10.1111/j.1365-2966.2006.10324.x}, \href
  {https://ui.adsabs.harvard.edu/abs/2006MNRAS.369..791F} {369, 791}

\bibitem[\protect\citeauthoryear{{Feast}, {Menzies}  \& {Whitelock}}{{Feast}
  et~al.}{2013}]{FeastLynga}
{Feast} M.~W.,  {Menzies} J.~W.,   {Whitelock} P.~A.,  2013, \mn@doi [\mnras]
  {10.1093/mnrasl/sls009}, \href
  {https://ui.adsabs.harvard.edu/abs/2013MNRAS.428L..36F} {428, L36}

\bibitem[\protect\citeauthoryear{{Fouesneau} et~al.,}{{Fouesneau}
  et~al.}{2022}]{Fouesneau2022}
{Fouesneau} M.,  et~al., 2022, \mn@doi [arXiv e-prints]
  {10.48550/arXiv.2206.05992}, \href
  {https://ui.adsabs.harvard.edu/abs/2022arXiv220605992F} {p. arXiv:2206.05992}

\bibitem[\protect\citeauthoryear{{Fraser}, {Hawley}  \& {Cook}}{{Fraser}
  et~al.}{2008}]{Fraser2008}
{Fraser} O.~J.,  {Hawley} S.~L.,   {Cook} K.~H.,  2008, \mn@doi [\aj]
  {10.1088/0004-6256/136/3/1242}, \href
  {https://ui.adsabs.harvard.edu/abs/2008AJ....136.1242F} {136, 1242}

\bibitem[\protect\citeauthoryear{{Fritz} et~al.,}{{Fritz}
  et~al.}{2011}]{Fritz2011}
{Fritz} T.~K.,  et~al., 2011, \mn@doi [\apj] {10.1088/0004-637X/737/2/73},
  \href {https://ui.adsabs.harvard.edu/abs/2011ApJ...737...73F} {737, 73}

\bibitem[\protect\citeauthoryear{{Gaia Collaboration} et~al.,}{{Gaia
  Collaboration} et~al.}{2016}]{Gaia1}
{Gaia Collaboration} et~al., 2016, \mn@doi [\aap]
  {10.1051/0004-6361/201629272}, \href
  {http://adsabs.harvard.edu/abs/2016A%26A...595A...1G} {595, A1}

\bibitem[\protect\citeauthoryear{{Gaia Collaboration} et~al.,}{{Gaia
  Collaboration} et~al.}{2021}]{GaiaEDR3}
{Gaia Collaboration} et~al., 2021, \mn@doi [\aap]
  {10.1051/0004-6361/202039657}, \href
  {https://ui.adsabs.harvard.edu/abs/2021A&A...649A...1G} {649, A1}

\bibitem[\protect\citeauthoryear{{Gaia Collaboration} et~al.,}{{Gaia
  Collaboration} et~al.}{2022}]{Creevey2022}
{Gaia Collaboration} et~al., 2022, arXiv e-prints, \href
  {https://ui.adsabs.harvard.edu/abs/2022arXiv220605870G} {p. arXiv:2206.05870}

\bibitem[\protect\citeauthoryear{{Gavel}, {Andrae}, {Fouesneau}, {Korn}  \&
  {Sordo}}{{Gavel} et~al.}{2021}]{Gavel2021}
{Gavel} A.,  {Andrae} R.,  {Fouesneau} M.,  {Korn} A.~J.,   {Sordo} R.,  2021,
  \mn@doi [\aap] {10.1051/0004-6361/202141589}, \href
  {https://ui.adsabs.harvard.edu/abs/2021A&A...656A..93G} {656, A93}

\bibitem[\protect\citeauthoryear{{Geller}, {Hurley}  \& {Mathieu}}{{Geller}
  et~al.}{2013}]{Geller2013}
{Geller} A.~M.,  {Hurley} J.~R.,   {Mathieu} R.~D.,  2013, \mn@doi [\aj]
  {10.1088/0004-6256/145/1/8}, \href
  {https://ui.adsabs.harvard.edu/abs/2013AJ....145....8G} {145, 8}

\bibitem[\protect\citeauthoryear{{Glass} \& {Evans}}{{Glass} \&
  {Evans}}{1981}]{GlassLloydEvans1981}
{Glass} I.~S.,  {Evans} T.~L.,  1981, \mn@doi [\nat] {10.1038/291303a0}, \href
  {https://ui.adsabs.harvard.edu/abs/1981Natur.291..303G} {291, 303}

\bibitem[\protect\citeauthoryear{{Gonneau} et~al.,}{{Gonneau}
  et~al.}{2016}]{Gonneau2016}
{Gonneau} A.,  et~al., 2016, \mn@doi [\aap] {10.1051/0004-6361/201526292},
  \href {https://ui.adsabs.harvard.edu/abs/2016A&A...589A..36G} {589, A36}

\bibitem[\protect\citeauthoryear{{Gonneau} et~al.,}{{Gonneau}
  et~al.}{2020}]{Gonneau2021}
{Gonneau} A.,  et~al., 2020, \mn@doi [\aap] {10.1051/0004-6361/201936825},
  \href {https://ui.adsabs.harvard.edu/abs/2020A&A...634A.133G} {634, A133}

\bibitem[\protect\citeauthoryear{{Grady}, {Belokurov}  \& {Evans}}{{Grady}
  et~al.}{2019}]{Grady2019}
{Grady} J.,  {Belokurov} V.,   {Evans} N.~W.,  2019, \mn@doi [\mnras]
  {10.1093/mnras/sty3284}, \href
  {https://ui.adsabs.harvard.edu/abs/2019MNRAS.483.3022G} {483, 3022}

\bibitem[\protect\citeauthoryear{{Grady}, {Belokurov}  \& {Evans}}{{Grady}
  et~al.}{2020}]{Grady2020}
{Grady} J.,  {Belokurov} V.,   {Evans} N.~W.,  2020, \mn@doi [\mnras]
  {10.1093/mnras/stz3617}, \href
  {https://ui.adsabs.harvard.edu/abs/2020MNRAS.492.3128G} {492, 3128}

\bibitem[\protect\citeauthoryear{{Groenewegen}}{{Groenewegen}}{2004}]{Groenewegen2004}
{Groenewegen} M.~A.~T.,  2004, \mn@doi [\aap] {10.1051/0004-6361:20047098},
  \href {https://ui.adsabs.harvard.edu/abs/2004A&A...425..595G} {425, 595}

\bibitem[\protect\citeauthoryear{{Groenewegen} \& {Sloan}}{{Groenewegen} \&
  {Sloan}}{2018}]{Groenewegen2018}
{Groenewegen} M.~A.~T.,  {Sloan} G.~C.,  2018, \mn@doi [\aap]
  {10.1051/0004-6361/201731089}, \href
  {https://ui.adsabs.harvard.edu/abs/2018A&A...609A.114G} {609, A114}

\bibitem[\protect\citeauthoryear{{Grondin}, {Webb}, {Leigh}, {Speagle}  \&
  {Khalifeh}}{{Grondin} et~al.}{2023}]{Grondin2022}
{Grondin} S.~M.,  {Webb} J.~J.,  {Leigh} N. W.~C.,  {Speagle} J.~S.,
  {Khalifeh} R.~J.,  2023, \mn@doi [\mnras] {10.1093/mnras/stac3367}, \href
  {https://ui.adsabs.harvard.edu/abs/2023MNRAS.518.4249G} {518, 4249}

\bibitem[\protect\citeauthoryear{{Harris}}{{Harris}}{2010}]{Harris2010}
{Harris} W.~E.,  2010, arXiv e-prints, \href
  {https://ui.adsabs.harvard.edu/abs/2010arXiv1012.3224H} {p. arXiv:1012.3224}

\bibitem[\protect\citeauthoryear{{Hasselquist} et~al.,}{{Hasselquist}
  et~al.}{2021}]{Hasselquist2021}
{Hasselquist} S.,  et~al., 2021, \mn@doi [\apj] {10.3847/1538-4357/ac25f9},
  \href {https://ui.adsabs.harvard.edu/abs/2021ApJ...923..172H} {923, 172}

\bibitem[\protect\citeauthoryear{{Herwig}}{{Herwig}}{2005}]{Herwig2005}
{Herwig} F.,  2005, \mn@doi [\araa] {10.1146/annurev.astro.43.072103.150600},
  \href {https://ui.adsabs.harvard.edu/abs/2005ARA&A..43..435H} {43, 435}

\bibitem[\protect\citeauthoryear{{H{\"o}fner} \& {Olofsson}}{{H{\"o}fner} \&
  {Olofsson}}{2018}]{Hofner2018}
{H{\"o}fner} S.,  {Olofsson} H.,  2018, \mn@doi [\aapr]
  {10.1007/s00159-017-0106-5}, \href
  {https://ui.adsabs.harvard.edu/abs/2018A&ARv..26....1H} {26, 1}

\bibitem[\protect\citeauthoryear{{Holl} et~al.,}{{Holl}
  et~al.}{2018}]{Holl2018}
{Holl} B.,  et~al., 2018, \mn@doi [\aap] {10.1051/0004-6361/201832892}, \href
  {https://ui.adsabs.harvard.edu/abs/2018A&A...618A..30H} {618, A30}

\bibitem[\protect\citeauthoryear{{Horta} et~al.,}{{Horta}
  et~al.}{2021a}]{Horta}
{Horta} D.,  et~al., 2021a, \mn@doi [\mnras] {10.1093/mnras/staa2987}, \href
  {https://ui.adsabs.harvard.edu/abs/2021MNRAS.500.1385H} {500, 1385}

\bibitem[\protect\citeauthoryear{{Horta} et~al.,}{{Horta}
  et~al.}{2021b}]{Horta2021}
{Horta} D.,  et~al., 2021b, \mn@doi [\mnras] {10.1093/mnras/staa3598}, \href
  {https://ui.adsabs.harvard.edu/abs/2021MNRAS.500.5462H} {500, 5462}

\bibitem[\protect\citeauthoryear{{Huang} et~al.,}{{Huang}
  et~al.}{2018}]{Huang2018}
{Huang} C.~D.,  et~al., 2018, \mn@doi [\apj] {10.3847/1538-4357/aab6b3}, \href
  {https://ui.adsabs.harvard.edu/abs/2018ApJ...857...67H} {857, 67}

\bibitem[\protect\citeauthoryear{{Huang} et~al.,}{{Huang}
  et~al.}{2020}]{Huang2020}
{Huang} C.~D.,  et~al., 2020, \mn@doi [\apj] {10.3847/1538-4357/ab5dbd}, \href
  {https://ui.adsabs.harvard.edu/abs/2020ApJ...889....5H} {889, 5}

\bibitem[\protect\citeauthoryear{{Hunter}}{{Hunter}}{2007}]{matplotlib}
{Hunter} J.~D.,  2007, Computing in Science Engineering, 9, 90

\bibitem[\protect\citeauthoryear{{Ishihara}, {Kaneda}, {Onaka}, {Ita},
  {Matsuura}  \& {Matsunaga}}{{Ishihara} et~al.}{2011}]{Ishihara2011}
{Ishihara} D.,  {Kaneda} H.,  {Onaka} T.,  {Ita} Y.,  {Matsuura} M.,
  {Matsunaga} N.,  2011, \mn@doi [\aap] {10.1051/0004-6361/201117626}, \href
  {https://ui.adsabs.harvard.edu/abs/2011A&A...534A..79I} {534, A79}

\bibitem[\protect\citeauthoryear{{Ita} \& {Matsunaga}}{{Ita} \&
  {Matsunaga}}{2011}]{Ita2011}
{Ita} Y.,  {Matsunaga} N.,  2011, \mn@doi [\mnras]
  {10.1111/j.1365-2966.2010.18056.x}, \href
  {https://ui.adsabs.harvard.edu/abs/2011MNRAS.412.2345I} {412, 2345}

\bibitem[\protect\citeauthoryear{{Ita} et~al.,}{{Ita} et~al.}{2004}]{Ita2004}
{Ita} Y.,  et~al., 2004, \mn@doi [\mnras] {10.1111/j.1365-2966.2004.07257.x},
  \href {https://ui.adsabs.harvard.edu/abs/2004MNRAS.347..720I} {347, 720}

\bibitem[\protect\citeauthoryear{{Iwanek}, {Soszy{\'n}ski}  \&
  {Koz{\l}owski}}{{Iwanek} et~al.}{2021}]{Iwanek2021b}
{Iwanek} P.,  {Soszy{\'n}ski} I.,   {Koz{\l}owski} S.,  2021, \mn@doi [\apj]
  {10.3847/1538-4357/ac10c5}, \href
  {https://ui.adsabs.harvard.edu/abs/2021ApJ...919...99I} {919, 99}

\bibitem[\protect\citeauthoryear{{Jadhav} \& {Subramaniam}}{{Jadhav} \&
  {Subramaniam}}{2021}]{Jadhav}
{Jadhav} V.~V.,  {Subramaniam} A.,  2021, \mn@doi [\mnras]
  {10.1093/mnras/stab2264}, \href
  {https://ui.adsabs.harvard.edu/abs/2021MNRAS.507.1699J} {507, 1699}

\bibitem[\protect\citeauthoryear{{Karakas}}{{Karakas}}{2014}]{Karakas2014b}
{Karakas} A.~I.,  2014, \mn@doi [\mnras] {10.1093/mnras/stu1727}, \href
  {https://ui.adsabs.harvard.edu/abs/2014MNRAS.445..347K} {445, 347}

\bibitem[\protect\citeauthoryear{{Karakas} \& {Lattanzio}}{{Karakas} \&
  {Lattanzio}}{2014}]{Karakas2014}
{Karakas} A.~I.,  {Lattanzio} J.~C.,  2014, \mn@doi [\pasa]
  {10.1017/pasa.2014.21}, \href
  {https://ui.adsabs.harvard.edu/abs/2014PASA...31...30K} {31, e030}

\bibitem[\protect\citeauthoryear{{Kastner}, {Thorndike}, {Romanczyk},
  {Buchanan}, {Hrivnak}, {Sahai}  \& {Egan}}{{Kastner}
  et~al.}{2008}]{Kastner2008}
{Kastner} J.~H.,  {Thorndike} S.~L.,  {Romanczyk} P.~A.,  {Buchanan} C.~L.,
  {Hrivnak} B.~J.,  {Sahai} R.,   {Egan} M.,  2008, \mn@doi [\aj]
  {10.1088/0004-6256/136/3/1221}, \href
  {https://ui.adsabs.harvard.edu/abs/2008AJ....136.1221K} {136, 1221}

\bibitem[\protect\citeauthoryear{Kim, Telea, Trager  \& Roerdink}{Kim
  et~al.}{2022}]{Kim2022}
Kim Y.,  Telea A.~C.,  Trager S.~C.,   Roerdink J.~B.,  2022, \mn@doi
  [Information Visualization] {10.1177/14738716221086589}, 21, 197

\bibitem[\protect\citeauthoryear{{Knigge}, {Leigh}  \& {Sills}}{{Knigge}
  et~al.}{2009}]{Knigge}
{Knigge} C.,  {Leigh} N.,   {Sills} A.,  2009, \mn@doi [\nat]
  {10.1038/nature07635}, \href
  {https://ui.adsabs.harvard.edu/abs/2009Natur.457..288K} {457, 288}

\bibitem[\protect\citeauthoryear{{Kobayashi}, {Karakas}  \&
  {Lugaro}}{{Kobayashi} et~al.}{2020}]{Kobayashi2020}
{Kobayashi} C.,  {Karakas} A.~I.,   {Lugaro} M.,  2020, \mn@doi [\apj]
  {10.3847/1538-4357/abae65}, \href
  {https://ui.adsabs.harvard.edu/abs/2020ApJ...900..179K} {900, 179}

\bibitem[\protect\citeauthoryear{{Koch}, {McWilliam}, {Preston}  \&
  {Thompson}}{{Koch} et~al.}{2016}]{Koch2016}
{Koch} A.,  {McWilliam} A.,  {Preston} G.~W.,   {Thompson} I.~B.,  2016,
  \mn@doi [\aap] {10.1051/0004-6361/201527413}, \href
  {https://ui.adsabs.harvard.edu/abs/2016A&A...587A.124K} {587, A124}

\bibitem[\protect\citeauthoryear{{Koposov} \& {Bartunov}}{{Koposov} \&
  {Bartunov}}{2006}]{q3c}
{Koposov} S.,  {Bartunov} O.,  2006, in {Gabriel} C.,  {Arviset} C.,  {Ponz}
  D.,   {Enrique} S.,  eds,  Astronomical Society of the Pacific Conference
  Series Vol. 351, Astronomical Data Analysis Software and Systems XV. p.~735

\bibitem[\protect\citeauthoryear{{Kraemer}, {Sloan}, {Price}  \&
  {Walker}}{{Kraemer} et~al.}{2002}]{Kraemer2002}
{Kraemer} K.~E.,  {Sloan} G.~C.,  {Price} S.~D.,   {Walker} H.~J.,  2002,
  \mn@doi [\apjs] {10.1086/339708}, \href
  {https://ui.adsabs.harvard.edu/abs/2002ApJS..140..389K} {140, 389}

\bibitem[\protect\citeauthoryear{{Kruijssen} et~al.,}{{Kruijssen}
  et~al.}{2020}]{Kruijssen}
{Kruijssen} J.~M.~D.,  et~al., 2020, \mn@doi [\mnras] {10.1093/mnras/staa2452},
  \href {https://ui.adsabs.harvard.edu/abs/2020MNRAS.498.2472K} {498, 2472}

\bibitem[\protect\citeauthoryear{{Lan{\c{c}}on} \& {Mouhcine}}{{Lan{\c{c}}on}
  \& {Mouhcine}}{2002}]{Lancon2002}
{Lan{\c{c}}on} A.,  {Mouhcine} M.,  2002, \mn@doi [\aap]
  {10.1051/0004-6361:20020585}, \href
  {https://ui.adsabs.harvard.edu/abs/2002A&A...393..167L} {393, 167}

\bibitem[\protect\citeauthoryear{{Lan{\c{c}}on} \& {Wood}}{{Lan{\c{c}}on} \&
  {Wood}}{2000}]{Lancon2000}
{Lan{\c{c}}on} A.,  {Wood} P.~R.,  2000, \mn@doi [\aaps] {10.1051/aas:2000269},
  \href {https://ui.adsabs.harvard.edu/abs/2000A&AS..146..217L} {146, 217}

\bibitem[\protect\citeauthoryear{{Lebzelter}, {Mowlavi}, {Marigo},
  {Pastorelli}, {Trabucchi}, {Wood}  \& {Lecoeur-Ta{\"\i}bi}}{{Lebzelter}
  et~al.}{2018}]{Lebzelter2018}
{Lebzelter} T.,  {Mowlavi} N.,  {Marigo} P.,  {Pastorelli} G.,  {Trabucchi} M.,
   {Wood} P.~R.,   {Lecoeur-Ta{\"\i}bi} I.,  2018, \mn@doi [\aap]
  {10.1051/0004-6361/201833615}, \href
  {https://ui.adsabs.harvard.edu/abs/2018A&A...616L..13L} {616, L13}

\bibitem[\protect\citeauthoryear{{Lebzelter} et~al.,}{{Lebzelter}
  et~al.}{2022}]{Lebzelter2022}
{Lebzelter} T.,  et~al., 2022, arXiv e-prints, \href
  {https://ui.adsabs.harvard.edu/abs/2022arXiv220605745L} {p. arXiv:2206.05745}

\bibitem[\protect\citeauthoryear{{Leiner} \& {Geller}}{{Leiner} \&
  {Geller}}{2021}]{LeinerGeller}
{Leiner} E.~M.,  {Geller} A.,  2021, \mn@doi [\apj] {10.3847/1538-4357/abd7e9},
  \href {https://ui.adsabs.harvard.edu/abs/2021ApJ...908..229L} {908, 229}

\bibitem[\protect\citeauthoryear{{Lewis}, {Pihlstr{\"o}m}, {Sjouwerman},
  {Stroh}, {Morris}  \& {BAaDE Collaboration}}{{Lewis}
  et~al.}{2020a}]{Lewis2020a}
{Lewis} M.~O.,  {Pihlstr{\"o}m} Y.~M.,  {Sjouwerman} L.~O.,  {Stroh} M.~C.,
  {Morris} M.~R.,   {BAaDE Collaboration} 2020a, \mn@doi [\apj]
  {10.3847/1538-4357/ab7920}, \href
  {https://ui.adsabs.harvard.edu/abs/2020ApJ...892...52L} {892, 52}

\bibitem[\protect\citeauthoryear{{Lewis}, {Pihlstr{\"o}m}, {Sjouwerman}  \&
  {Quiroga-Nu{\~n}ez}}{{Lewis} et~al.}{2020b}]{Lewis2020b}
{Lewis} M.~O.,  {Pihlstr{\"o}m} Y.~M.,  {Sjouwerman} L.~O.,
  {Quiroga-Nu{\~n}ez} L.~H.,  2020b, \mn@doi [\apj] {10.3847/1538-4357/abaf46},
  \href {https://ui.adsabs.harvard.edu/abs/2020ApJ...901...98L} {901, 98}

\bibitem[\protect\citeauthoryear{{Lian}, {Zhu}, {Kong}  \& {He}}{{Lian}
  et~al.}{2014}]{Lian2014}
{Lian} J.,  {Zhu} Q.,  {Kong} X.,   {He} J.,  2014, \mn@doi [\aap]
  {10.1051/0004-6361/201322818}, \href
  {https://ui.adsabs.harvard.edu/abs/2014A&A...564A..84L} {564, A84}

\bibitem[\protect\citeauthoryear{{Liu}, {Bailer-Jones}, {Sordo}, {Vallenari},
  {Borrachero}, {Luri}  \& {Sartoretti}}{{Liu} et~al.}{2012}]{Liu2012}
{Liu} C.,  {Bailer-Jones} C.~A.~L.,  {Sordo} R.,  {Vallenari} A.,  {Borrachero}
  R.,  {Luri} X.,   {Sartoretti} P.,  2012, \mn@doi [\mnras]
  {10.1111/j.1365-2966.2012.21797.x}, \href
  {https://ui.adsabs.harvard.edu/abs/2012MNRAS.426.2463L} {426, 2463}

\bibitem[\protect\citeauthoryear{{Lloyd Evans}}{{Lloyd
  Evans}}{2010}]{LloydEvans2010}
{Lloyd Evans} T.,  2010, \mn@doi [Journal of Astrophysics and Astronomy]
  {10.1007/s12036-010-0017-6}, \href
  {https://ui.adsabs.harvard.edu/abs/2010JApA...31..177L} {31, 177}

\bibitem[\protect\citeauthoryear{{L{\'o}pez-Corredoira}}{{L{\'o}pez-Corredoira}}{2017}]{LopezCorredoira2017}
{L{\'o}pez-Corredoira} M.,  2017, \mn@doi [\apj] {10.3847/1538-4357/836/2/218},
  \href {https://ui.adsabs.harvard.edu/abs/2017ApJ...836..218L} {836, 218}

\bibitem[\protect\citeauthoryear{{Lucey} et~al.,}{{Lucey}
  et~al.}{2022}]{Lucey2022}
{Lucey} M.,  et~al., 2022, arXiv e-prints, \href
  {https://ui.adsabs.harvard.edu/abs/2022arXiv220608299L} {p. arXiv:2206.08299}

\bibitem[\protect\citeauthoryear{{MacConnell}}{{MacConnell}}{1988}]{MacConnell1988}
{MacConnell} D.~J.,  1988, \mn@doi [\aj] {10.1086/114813}, \href
  {https://ui.adsabs.harvard.edu/abs/1988AJ.....96..354M} {96, 354}

\bibitem[\protect\citeauthoryear{{Marigo} et~al.,}{{Marigo}
  et~al.}{2022}]{Marigo2022}
{Marigo} P.,  et~al., 2022, \mn@doi [\apjs] {10.3847/1538-4365/ac374a}, \href
  {https://ui.adsabs.harvard.edu/abs/2022ApJS..258...43M} {258, 43}

\bibitem[\protect\citeauthoryear{Matsunaga, Menzies, Feast, Whitelock, Onozato,
  Barway  \& Aydi}{Matsunaga et~al.}{2017}]{Matsunaga2017}
Matsunaga N.,  Menzies J.~W.,  Feast M.~W.,  Whitelock P.~A.,  Onozato H.,
  Barway S.,   Aydi E.,  2017, \mn@doi [\mnras] {10.1093/mnras/stx1213}, 469,
  4949

\bibitem[\protect\citeauthoryear{{McInnes}, {Healy}  \& {Melville}}{{McInnes}
  et~al.}{2018}]{McInnes2018}
{McInnes} L.,  {Healy} J.,   {Melville} J.,  2018, arXiv e-prints, \href
  {https://ui.adsabs.harvard.edu/abs/2018arXiv180203426M} {p. arXiv:1802.03426}

\bibitem[\protect\citeauthoryear{{McKinney}}{{McKinney}}{2010}]{pandas1}
{McKinney} W.,  2010, in {S}t\'efan van~der {W}alt {J}arrod {M}illman eds,
  {P}roceedings of the 9th {P}ython in {S}cience {C}onference. pp 56 -- 61,
  \mn@doi{10.25080/Majora-92bf1922-00a}

\bibitem[\protect\citeauthoryear{{Milone} et~al.,}{{Milone}
  et~al.}{2012}]{Milone2012}
{Milone} A.~P.,  et~al., 2012, \mn@doi [\aap] {10.1051/0004-6361/201016384},
  \href {https://ui.adsabs.harvard.edu/abs/2012A&A...540A..16M} {540, A16}

\bibitem[\protect\citeauthoryear{{Miszalski}, {Miko{\l}ajewska}  \&
  {Udalski}}{{Miszalski} et~al.}{2013}]{Miszalski2013}
{Miszalski} B.,  {Miko{\l}ajewska} J.,   {Udalski} A.,  2013, \mn@doi [\mnras]
  {10.1093/mnras/stt673}, \href
  {https://ui.adsabs.harvard.edu/abs/2013MNRAS.432.3186M} {432, 3186}

\bibitem[\protect\citeauthoryear{{Montegriffo} et~al.,}{{Montegriffo}
  et~al.}{2022}]{Montegriffo2022}
{Montegriffo} P.,  et~al., 2022, arXiv e-prints, \href
  {https://ui.adsabs.harvard.edu/abs/2022arXiv220606205M} {p. arXiv:2206.06205}

\bibitem[\protect\citeauthoryear{{Mowlavi} et~al.,}{{Mowlavi}
  et~al.}{2018}]{Mowlavi2018}
{Mowlavi} N.,  et~al., 2018, \mn@doi [\aap] {10.1051/0004-6361/201833366},
  \href {https://ui.adsabs.harvard.edu/abs/2018A%26A...618A..58M} {618, A58}

\bibitem[\protect\citeauthoryear{{M{\"u}rset} \& {Schmid}}{{M{\"u}rset} \&
  {Schmid}}{1999}]{Murset1999}
{M{\"u}rset} U.,  {Schmid} H.~M.,  1999, \mn@doi [\aaps] {10.1051/aas:1999105},
  \href {https://ui.adsabs.harvard.edu/abs/1999A&AS..137..473M} {137, 473}

\bibitem[\protect\citeauthoryear{{Nassau} \& {Velghe}}{{Nassau} \&
  {Velghe}}{1964}]{Nassau1964}
{Nassau} J.~J.,  {Velghe} A.~G.,  1964, \mn@doi [\apj] {10.1086/147745}, \href
  {https://ui.adsabs.harvard.edu/abs/1964ApJ...139..190N} {139, 190}

\bibitem[\protect\citeauthoryear{{Nataf}}{{Nataf}}{2016}]{Nataf2016}
{Nataf} D.~M.,  2016, \mn@doi [\pasa] {10.1017/pasa.2015.38}, \href
  {https://ui.adsabs.harvard.edu/abs/2016PASA...33...23N} {33, e023}

\bibitem[\protect\citeauthoryear{{Ng}}{{Ng}}{1997}]{Ng1997}
{Ng} Y.~K.,  1997, \aap, \href
  {https://ui.adsabs.harvard.edu/abs/1997A&A...328..211N} {328, 211}

\bibitem[\protect\citeauthoryear{{Ng}}{{Ng}}{1998}]{Ng1998}
{Ng} Y.~K.,  1998, \aap, \href
  {https://ui.adsabs.harvard.edu/abs/1998A&A...338..435N} {338, 435}

\bibitem[\protect\citeauthoryear{{Nikutta}, {Hunt-Walker}, {Nenkova},
  {Ivezi{\'c}}  \& {Elitzur}}{{Nikutta} et~al.}{2014}]{Nikutta2014}
{Nikutta} R.,  {Hunt-Walker} N.,  {Nenkova} M.,  {Ivezi{\'c}} {\v{Z}}.,
  {Elitzur} M.,  2014, \mn@doi [\mnras] {10.1093/mnras/stu1087}, \href
  {https://ui.adsabs.harvard.edu/abs/2014MNRAS.442.3361N} {442, 3361}

\bibitem[\protect\citeauthoryear{{Nikzat} et~al.,}{{Nikzat}
  et~al.}{2022}]{Nikzat2022}
{Nikzat} F.,  et~al., 2022, \mn@doi [\aap] {10.1051/0004-6361/202141805}, \href
  {https://ui.adsabs.harvard.edu/abs/2022A&A...660A..35N} {660, A35}

\bibitem[\protect\citeauthoryear{{Nishiyama}, {Tamura}, {Hatano}, {Kato},
  {Tanab{\'e}}, {Sugitani}  \& {Nagata}}{{Nishiyama}
  et~al.}{2009}]{Nishiyama2009}
{Nishiyama} S.,  {Tamura} M.,  {Hatano} H.,  {Kato} D.,  {Tanab{\'e}} T.,
  {Sugitani} K.,   {Nagata} T.,  2009, \mn@doi [\apj]
  {10.1088/0004-637X/696/2/1407}, \href
  {https://ui.adsabs.harvard.edu/abs/2009ApJ...696.1407N} {696, 1407}

\bibitem[\protect\citeauthoryear{{Olnon} et~al.,}{{Olnon}
  et~al.}{1986}]{Olnon1986}
{Olnon} F.~M.,  et~al., 1986, \aaps, \href
  {https://ui.adsabs.harvard.edu/abs/1986A&AS...65..607O} {65, 607}

\bibitem[\protect\citeauthoryear{{Pietrzy{\'n}ski} et~al.,}{{Pietrzy{\'n}ski}
  et~al.}{2019}]{Pietrzynski2019}
{Pietrzy{\'n}ski} G.,  et~al., 2019, \mn@doi [\nat]
  {10.1038/s41586-019-0999-4}, \href
  {https://ui.adsabs.harvard.edu/abs/2019Natur.567..200P} {567, 200}

\bibitem[\protect\citeauthoryear{Poli{\v c}ar, Stra{\v z}ar  \& Zupan}{Poli{\v
  c}ar et~al.}{2019}]{openTSNE}
Poli{\v c}ar P.~G.,  Stra{\v z}ar M.,   Zupan B.,  2019, \mn@doi [bioRxiv]
  {10.1101/731877}

\bibitem[\protect\citeauthoryear{{Price-Whelan} et~al.,}{{Price-Whelan}
  et~al.}{2018}]{astropy:2018}
{Price-Whelan} A.~M.,  et~al., 2018, \mn@doi [\aj] {10.3847/1538-3881/aabc4f},
  \href {https://ui.adsabs.harvard.edu/#abs/2018AJ....156..123T} {156, 123}

\bibitem[\protect\citeauthoryear{{Reid} \& {Goldston}}{{Reid} \&
  {Goldston}}{2002}]{Reid2002}
{Reid} M.~J.,  {Goldston} J.~E.,  2002, \mn@doi [\apj] {10.1086/338947}, \href
  {https://ui.adsabs.harvard.edu/abs/2002ApJ...568..931R} {568, 931}

\bibitem[\protect\citeauthoryear{{Reis}, {Rotman}, {Poznanski}, {Prochaska}  \&
  {Wolf}}{{Reis} et~al.}{2019}]{Reis2019}
{Reis} I.,  {Rotman} M.,  {Poznanski} D.,  {Prochaska} J.~X.,   {Wolf} L.,
  2019, arXiv e-prints, \href
  {https://ui.adsabs.harvard.edu/abs/2019arXiv191106823R} {p. arXiv:1911.06823}

\bibitem[\protect\citeauthoryear{{Riebel}, {Meixner}, {Fraser}, {Srinivasan},
  {Cook}  \& {Vijh}}{{Riebel} et~al.}{2010}]{Riebel2010}
{Riebel} D.,  {Meixner} M.,  {Fraser} O.,  {Srinivasan} S.,  {Cook} K.,
  {Vijh} U.,  2010, \mn@doi [\apj] {10.1088/0004-637X/723/2/1195}, \href
  {https://ui.adsabs.harvard.edu/abs/2010ApJ...723.1195R} {723, 1195}

\bibitem[\protect\citeauthoryear{{Riello} et~al.,}{{Riello}
  et~al.}{2021}]{Riello2021}
{Riello} M.,  et~al., 2021, \mn@doi [\aap] {10.1051/0004-6361/202039587}, \href
  {https://ui.adsabs.harvard.edu/abs/2021A&A...649A...3R} {649, A3}

\bibitem[\protect\citeauthoryear{{Rimoldini} et~al.,}{{Rimoldini}
  et~al.}{2019}]{Rimoldini2019}
{Rimoldini} L.,  et~al., 2019, \mn@doi [\aap] {10.1051/0004-6361/201834616},
  \href {https://ui.adsabs.harvard.edu/abs/2019A&A...625A..97R} {625, A97}

\bibitem[\protect\citeauthoryear{{Rix} et~al.,}{{Rix} et~al.}{2022}]{Rix2022}
{Rix} H.-W.,  et~al., 2022, \mn@doi [\apj] {10.3847/1538-4357/ac9e01}, \href
  {https://ui.adsabs.harvard.edu/abs/2022ApJ...941...45R} {941, 45}

\bibitem[\protect\citeauthoryear{{Rodrigo} \& {Solano}}{{Rodrigo} \&
  {Solano}}{2020}]{SVO2}
{Rodrigo} C.,  {Solano} E.,  2020, in Contributions to the XIV.0 Scientific
  Meeting (virtual) of the Spanish Astronomical Society. p.~182

\bibitem[\protect\citeauthoryear{{Rodrigo}, {Solano}  \& {Bayo}}{{Rodrigo}
  et~al.}{2012}]{SVO1}
{Rodrigo} C.,  {Solano} E.,   {Bayo} A.,  2012, {SVO Filter Profile Service
  Version 1.0}, IVOA Working Draft 15 October 2012,
  \mn@doi{10.5479/ADS/bib/2012ivoa.rept.1015R}

\bibitem[\protect\citeauthoryear{Sanders, Smith, Evans  \& Lucas}{Sanders
  et~al.}{2019}]{Sanders2019a}
Sanders J.~L.,  Smith L.,  Evans N.~W.,   Lucas P.,  2019, \mn@doi [\mnras]
  {10.1093/mnras/stz1630}, 487, 5188

\bibitem[\protect\citeauthoryear{{Sanders}, {Smith},
  {Gonz{\'a}lez-Fern{\'a}ndez}, {Lucas}  \& {Minniti}}{{Sanders}
  et~al.}{2022a}]{Sanders2022_Ext}
{Sanders} J.~L.,  {Smith} L.,  {Gonz{\'a}lez-Fern{\'a}ndez} C.,  {Lucas} P.,
  {Minniti} D.,  2022a, \mn@doi [\mnras] {10.1093/mnras/stac1367}, \href
  {https://ui.adsabs.harvard.edu/abs/2022MNRAS.514.2407S} {514, 2407}

\bibitem[\protect\citeauthoryear{{Sanders}, {Matsunaga}, {Kawata}, {Smith},
  {Minniti}  \& {Lucas}}{{Sanders} et~al.}{2022b}]{Sanders2022}
{Sanders} J.~L.,  {Matsunaga} N.,  {Kawata} D.,  {Smith} L.~C.,  {Minniti} D.,
   {Lucas} P.~W.,  2022b, \mn@doi [\mnras] {10.1093/mnras/stac2274}, \href
  {https://ui.adsabs.harvard.edu/abs/2022MNRAS.517..257S} {517, 257}

\bibitem[\protect\citeauthoryear{{Schlegel}, {Finkbeiner}  \&
  {Davis}}{{Schlegel} et~al.}{1998}]{SFD}
{Schlegel} D.~J.,  {Finkbeiner} D.~P.,   {Davis} M.,  1998, \mn@doi [\apj]
  {10.1086/305772}, \href
  {https://ui.adsabs.harvard.edu/abs/1998ApJ...500..525S} {500, 525}

\bibitem[\protect\citeauthoryear{{Secchi}}{{Secchi}}{1868}]{Secchi1868}
{Secchi} A.,  1868, \mnras, \href
  {https://ui.adsabs.harvard.edu/abs/1868MNRAS..28..196S} {28, 196}

\bibitem[\protect\citeauthoryear{{Semczuk}, {Dehnen}, {Sch{\"o}nrich}  \&
  {Athanassoula}}{{Semczuk} et~al.}{2022}]{Semczuk2022}
{Semczuk} M.,  {Dehnen} W.,  {Sch{\"o}nrich} R.,   {Athanassoula} E.,  2022,
  \mn@doi [\mnras] {10.1093/mnras/stac3085}, \href
  {https://ui.adsabs.harvard.edu/abs/2022MNRAS.517.6060S} {517, 6060}

\bibitem[\protect\citeauthoryear{{Sharpless}}{{Sharpless}}{1956}]{Sharpless1956}
{Sharpless} S.,  1956, \mn@doi [\apj] {10.1086/146229}, \href
  {https://ui.adsabs.harvard.edu/abs/1956ApJ...124..342S} {124, 342}

\bibitem[\protect\citeauthoryear{{Simion}, {Belokurov}, {Irwin}, {Koposov},
  {Gonzalez-Fernandez}, {Robin}, {Shen}  \& {Li}}{{Simion}
  et~al.}{2017}]{Simion2017}
{Simion} I.~T.,  {Belokurov} V.,  {Irwin} M.,  {Koposov} S.~E.,
  {Gonzalez-Fernandez} C.,  {Robin} A.~C.,  {Shen} J.,   {Li} Z.~Y.,  2017,
  \mn@doi [\mnras] {10.1093/mnras/stx1832}, \href
  {https://ui.adsabs.harvard.edu/abs/2017MNRAS.471.4323S} {471, 4323}

\bibitem[\protect\citeauthoryear{{Soszy{\'n}ski} et~al.,}{{Soszy{\'n}ski}
  et~al.}{2009}]{Soszynski2009}
{Soszy{\'n}ski} I.,  et~al., 2009, \actaa, \href
  {https://ui.adsabs.harvard.edu/abs/2009AcA....59..239S} {59, 239}

\bibitem[\protect\citeauthoryear{{Suh} \& {Hong}}{{Suh} \&
  {Hong}}{2017}]{SuhHong2017}
{Suh} K.-W.,  {Hong} J.,  2017, \mn@doi [Journal of Korean Astronomical
  Society] {10.5303/JKAS.2017.50.4.131}, \href
  {https://ui.adsabs.harvard.edu/abs/2017JKAS...50..131S} {50, 131}

\bibitem[\protect\citeauthoryear{{Trabucchi} \& {Mowlavi}}{{Trabucchi} \&
  {Mowlavi}}{2022}]{Trabucchi2022}
{Trabucchi} M.,  {Mowlavi} N.,  2022, \mn@doi [\aap]
  {10.1051/0004-6361/202142853}, \href
  {https://ui.adsabs.harvard.edu/abs/2022A&A...658L...1T} {658, L1}

\bibitem[\protect\citeauthoryear{{Trabucchi}, {Wood}, {Montalb{\'a}n},
  {Marigo}, {Pastorelli}  \& {Girardi}}{{Trabucchi}
  et~al.}{2019}]{Trabucchi2019}
{Trabucchi} M.,  {Wood} P.~R.,  {Montalb{\'a}n} J.,  {Marigo} P.,  {Pastorelli}
  G.,   {Girardi} L.,  2019, \mn@doi [\mnras] {10.1093/mnras/sty2745}, \href
  {https://ui.adsabs.harvard.edu/abs/2019MNRAS.482..929T} {482, 929}

\bibitem[\protect\citeauthoryear{{Traven} et~al.,}{{Traven}
  et~al.}{2017}]{Traven2017}
{Traven} G.,  et~al., 2017, \mn@doi [\apjs] {10.3847/1538-4365/228/2/24}, \href
  {https://ui.adsabs.harvard.edu/abs/2017ApJS..228...24T} {228, 24}

\bibitem[\protect\citeauthoryear{{Van Eck} et~al.,}{{Van Eck}
  et~al.}{2017}]{VanEck2017}
{Van Eck} S.,  et~al., 2017, \mn@doi [\aap] {10.1051/0004-6361/201525886},
  \href {https://ui.adsabs.harvard.edu/abs/2017A&A...601A..10V} {601, A10}

\bibitem[\protect\citeauthoryear{{\VAN{Van}{van}{van} der Walt}, {Colbert}  \&
  {Varoquaux}}{{\VAN{Van}{van}{van} der Walt} et~al.}{2011}]{numpy}
{\VAN{Van}{van}{van} der Walt} S.,  {Colbert} S.~C.,   {Varoquaux} G.,  2011,
  Computing in Science Engineering, 13, 22

\bibitem[\protect\citeauthoryear{{Verro} et~al.,}{{Verro}
  et~al.}{2022}]{Verro2022}
{Verro} K.,  et~al., 2022, \mn@doi [\aap] {10.1051/0004-6361/202142388}, \href
  {https://ui.adsabs.harvard.edu/abs/2022A&A...660A..34V} {660, A34}

\bibitem[\protect\citeauthoryear{{Virtanen} et~al.,}{{Virtanen}
  et~al.}{2020}]{scipy}
{Virtanen} P.,  et~al., 2020, \mn@doi [Nature Methods]
  {https://doi.org/10.1038/s41592-019-0686-2}, \href {https://rdcu.be/b08Wh}
  {17, 261}

\bibitem[\protect\citeauthoryear{{Wallerstein} \& {Knapp}}{{Wallerstein} \&
  {Knapp}}{1998}]{Wallerstein1998}
{Wallerstein} G.,  {Knapp} G.~R.,  1998, \mn@doi [\araa]
  {10.1146/annurev.astro.36.1.369}, \href
  {https://ui.adsabs.harvard.edu/abs/1998ARA&A..36..369W} {36, 369}

\bibitem[\protect\citeauthoryear{Waskom et~al.,}{Waskom et~al.}{2017}]{seaborn}
Waskom M.,  et~al., 2017, seaborn: v0.8.1.
Zenodo, \mn@doi{10.5281/zenodo.883859}

\bibitem[\protect\citeauthoryear{{Wegg} \& {Gerhard}}{{Wegg} \&
  {Gerhard}}{2013}]{WeggGerhard2013}
{Wegg} C.,  {Gerhard} O.,  2013, \mn@doi [\mnras] {10.1093/mnras/stt1376},
  \href {https://ui.adsabs.harvard.edu/abs/2013MNRAS.435.1874W} {435, 1874}

\bibitem[\protect\citeauthoryear{{Wenger} et~al.,}{{Wenger}
  et~al.}{2000}]{Simbad}
{Wenger} M.,  et~al., 2000, \mn@doi [\aaps] {10.1051/aas:2000332}, \href
  {https://ui.adsabs.harvard.edu/abs/2000A&AS..143....9W} {143, 9}

\bibitem[\protect\citeauthoryear{{Whitehouse}, {Farihi}, {Green}, {Wilson}  \&
  {Subasavage}}{{Whitehouse} et~al.}{2018}]{Whitehouse2018}
{Whitehouse} L.~J.,  {Farihi} J.,  {Green} P.~J.,  {Wilson} T.~G.,
  {Subasavage} J.~P.,  2018, \mn@doi [\mnras] {10.1093/mnras/sty1622}, \href
  {https://ui.adsabs.harvard.edu/abs/2018MNRAS.479.3873W} {479, 3873}

\bibitem[\protect\citeauthoryear{{Witten} et~al.,}{{Witten}
  et~al.}{2022}]{Witten2022}
{Witten} C. E.~C.,  et~al., 2022, \mn@doi [\mnras] {10.1093/mnras/stac2273},
  \href {https://ui.adsabs.harvard.edu/abs/2022MNRAS.516.3254W} {516, 3254}

\bibitem[\protect\citeauthoryear{{Wood}}{{Wood}}{2000}]{Wood2000}
{Wood} P.~R.,  2000, \mn@doi [\pasa] {10.1071/AS00018}, \href
  {https://ui.adsabs.harvard.edu/abs/2000PASA...17...18W} {17, 18}

\bibitem[\protect\citeauthoryear{{Wyatt} \& {Cahn}}{{Wyatt} \&
  {Cahn}}{1983}]{WyattCahn1983}
{Wyatt} S.~P.,  {Cahn} J.~H.,  1983, \mn@doi [\apj] {10.1086/161527}, \href
  {https://ui.adsabs.harvard.edu/abs/1983ApJ...275..225W} {275, 225}

\bibitem[\protect\citeauthoryear{{Xylakis-Dornbusch}, {Christlieb}, {Lind}  \&
  {Nordlander}}{{Xylakis-Dornbusch} et~al.}{2022}]{Xylakis2022}
{Xylakis-Dornbusch} T.,  {Christlieb} N.,  {Lind} K.,   {Nordlander} T.,  2022,
  arXiv e-prints, \href {https://ui.adsabs.harvard.edu/abs/2022arXiv220802027X}
  {p. arXiv:2208.02027}

\bibitem[\protect\citeauthoryear{{Yao}, {Liu}, {Deng}, {de Grijs}  \&
  {Matsunaga}}{{Yao} et~al.}{2017}]{Yao2017}
{Yao} Y.,  {Liu} C.,  {Deng} L.,  {de Grijs} R.,   {Matsunaga} N.,  2017,
  \mn@doi [\apjs] {10.3847/1538-4365/aa88a9}, \href
  {https://ui.adsabs.harvard.edu/abs/2017ApJS..232...16Y} {232, 16}

\bibitem[\protect\citeauthoryear{{Yuan}, {He}, {Macri}, {Long}  \&
  {Huang}}{{Yuan} et~al.}{2017a}]{Yuan2017}
{Yuan} W.,  {He} S.,  {Macri} L.~M.,  {Long} J.,   {Huang} J.~Z.,  2017a,
  \mn@doi [\aj] {10.3847/1538-3881/aa63f1}, \href
  {https://ui.adsabs.harvard.edu/abs/2017AJ....153..170Y} {153, 170}

\bibitem[\protect\citeauthoryear{{Yuan}, {Macri}, {He}, {Huang}, {Kanbur}  \&
  {Ngeow}}{{Yuan} et~al.}{2017b}]{Yuan2018}
{Yuan} W.,  {Macri} L.~M.,  {He} S.,  {Huang} J.~Z.,  {Kanbur} S.~M.,   {Ngeow}
  C.-C.,  2017b, \mn@doi [\aj] {10.3847/1538-3881/aa86f1}, \href
  {https://ui.adsabs.harvard.edu/abs/2017AJ....154..149Y} {154, 149}

\bibitem[\protect\citeauthoryear{{Zoccali} et~al.,}{{Zoccali}
  et~al.}{2003}]{Zoccali2003}
{Zoccali} M.,  et~al., 2003, \mn@doi [\aap] {10.1051/0004-6361:20021604}, \href
  {https://ui.adsabs.harvard.edu/abs/2003A&A...399..931Z} {399, 931}

\bibitem[\protect\citeauthoryear{van~der Maaten \& Hinton}{van~der Maaten \&
  Hinton}{2008}]{tSNE}
van~der Maaten L.,  Hinton G.,  2008, Journal of Machine Learning Research, 9,
  2579

\makeatother
\end{thebibliography}


\appendix
\section{Classification of the entire Gaia DR3 long-period variable catalogue}\label{appendix::full}
\begin{figure*}
    \centering
    \includegraphics[width=\textwidth]{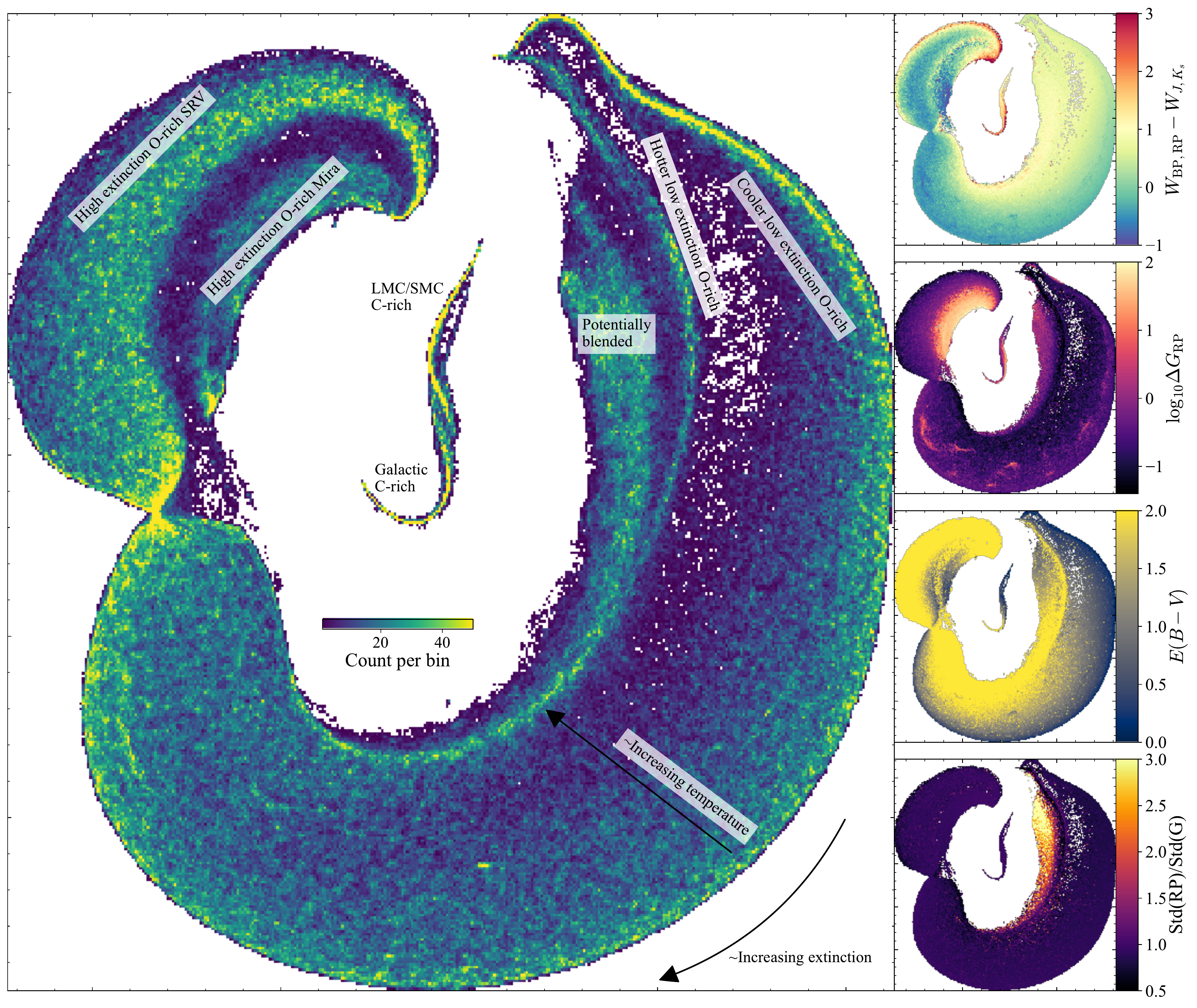}
    \caption{Two-dimensional UMAP projection of \emph{all} Gaia DR3 long-period variable candidates. The main panel shows the counts per bin whilst each of the right subpanels shows the same diagram coloured by different properties.}
    \label{fig:all_umap}
\end{figure*}
In the main body of the paper, we have focused on the high-amplitude variable stars in the Gaia DR3 long-period variable candidates catalogue. These are likely highly reliable but also contain the interesting Mira variable subset useful as a Local Group and cosmological distance and age tracer. In this appendix, we extend the analysis to the entire LPV catalogue of $1,205,121$ stars with Gaia BP/RP spectra. We run the same UMAP computation described in Section~\ref{sec:unsupervisedClassification} on the full dataset and display the results in Fig.~\ref{fig:all_umap}. As with the high-amplitude sample, we see two distinct regions -- a crescent of O-rich sources and a spur of C-rich sources. Again, the C-rich spur forms a near one-dimensional sequence corresponding primarily to variations in extinction (as the C-rich features are only weakly sensitive to effective temperature). However, unlike Fig.~\ref{fig:umap} we observe the spur is almost two overlaid one-dimensional sequences which we identify as due to the LMC/SMC sources and the Galactic sources respectively. The O-rich crescent is more structured than the C-rich spur due to the combination of extinction and effective temperature variation. If we consider the right part of the crescent, there are three overdense features. The right feature is composed of low extinction cooler stars often in the LMC/SMC whilst the middle sequence tends to be hotter stars without significant spectral features and with on average higher extinction. The left feature appears to be due to blended/contaminated sources as evidenced by their large ratio of the $G_\mathrm{RP}$ standard deviation to the equivalent in $G$. As we move clockwise around the crescent, the sources are typically higher extinction but also have the tendency to be cooler with more pronounced spectral features. The high-amplitude Mira variables sit on the right edge of the left part of the crescent. The directions in the UMAP diagram are awkward to map to physical dimensions but we have found moving across the crescent approximately maps into temperature variation whilst moving around the crescent maps into extinction variations. However, as evidenced by the diagram coloured by extinction, this isn't a perfect mapping.

\section{Comparison of amplitude measures for long-period variables}\label{section::amplitude_comparison}
\begin{figure*}
    \centering
    \includegraphics[width=\textwidth]{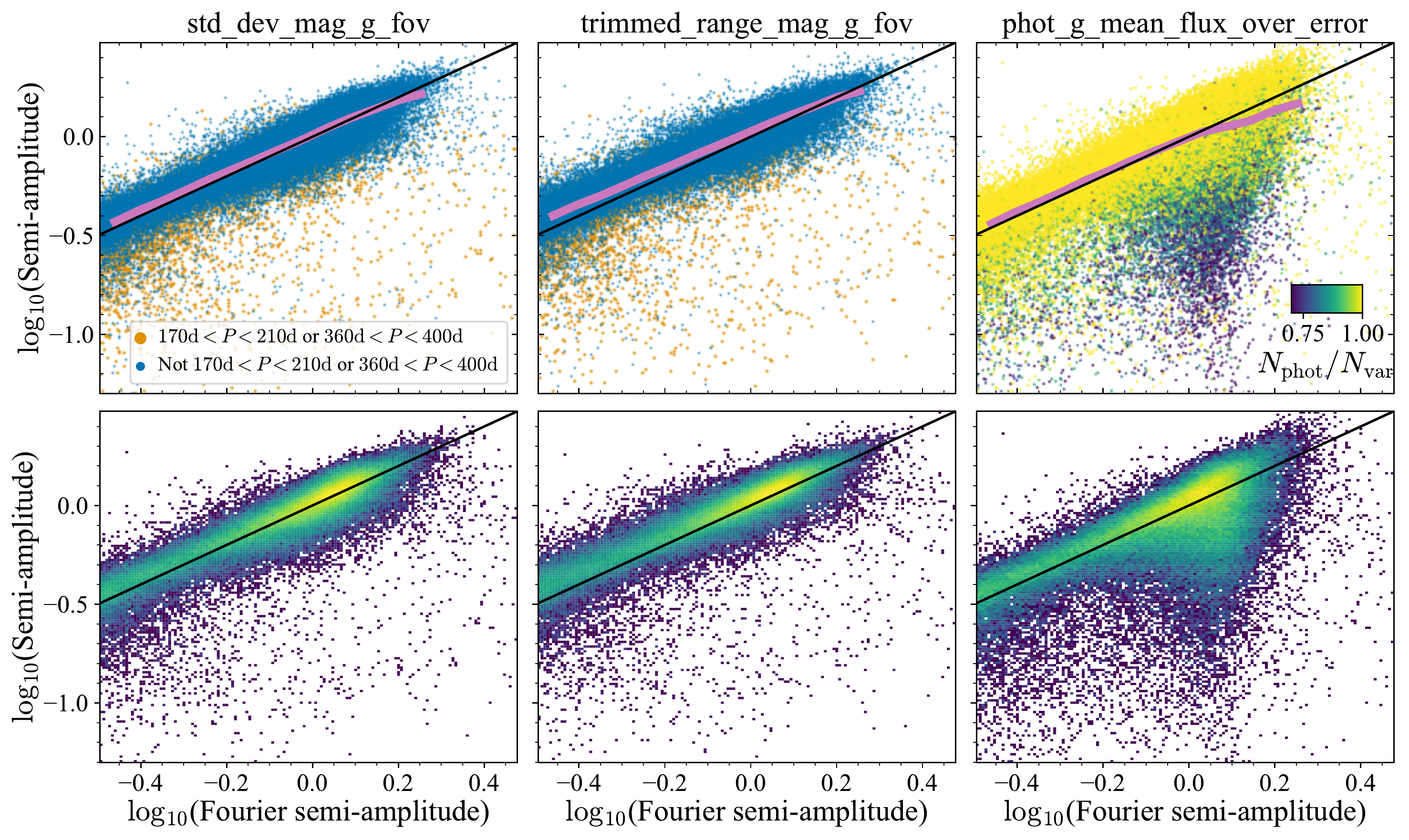}
    \caption{Comparison of amplitude measures for Gaia long-period variable stars. Every panel shows the semi-amplitude of a Fourier fit (denoted $\Delta G_\mathrm{Fourier}$ in the text) against one of the other amplitude measures: $\Delta G_\mathrm{std}$ from the standard deviation of the epoch photometry in the first column, $\Delta G_\mathrm{range}$ from the $95$th-$5$th percentile range in the second column and $\Delta G_\mathrm{phot}$ from the photometric uncertainties in the third column. Both rows show the same data -- the lower row is a logarithmically-coloured histogram. The top row is scatter plots coloured by whether the period is near an alias (orange) or not (blue) in the left two rows, and by the ratio of the number of measurements used in the mean photometry compared to the number used in the variable star epoch photometry processing. The solid pink line gives the median trend. The black line is a one-to-one relation. Note sources with periods near $190$ and $380$ day typically have overestimated Fourier amplitudes and/or underestimated amplitudes using the other methods as they only measure the scatter of the available data. If a low number of photometric points are used in the mean photometric pipeline, the amplitude estimated from the photometric uncertainties is typically smaller than the Fourier amplitude.}
    \label{fig:amp_measure}
\end{figure*}

Within the Gaia DR3 variable star catalogues, there are multiple measures of the variability amplitude. As highlighted by \cite{Belokurov2016}, the Gaia photometric uncertainties contain variability information. As they are computed as errors in the mean of the epoch photometric measurements, the semi-amplitude can be estimated as
\begin{equation}
    \Delta G_\mathrm{phot} = \frac{2.5\sqrt{2}}{\ln 10}\frac{\sqrt{\texttt{phot\_g\_n\_obs}}}{\texttt{phot\_g\_mean\_flux\_over\_error}}.
\end{equation}
For those Gaia sources classified as variable \citep{Holl2018,Rimoldini2019}, the standard deviation of the epoch photometry is reported as \texttt{std\_dev\_fov\_g} from which the semi-amplitude can be estimated as
\begin{equation}
    \Delta G_\mathrm{std} = \sqrt{2}\,\texttt{std\_dev\_fov\_g}.
\end{equation}
Furthermore, the $95$th-$5$th percentile, \texttt{trimmed\_range\_mag\_g\_fov}, is reported for these sources from which we can find
\begin{equation}
    \Delta G_\mathrm{range} = \frac{1}{2\cos(\pi/20)}\texttt{trimmed\_range\_mag\_g\_fov}.
\end{equation}
Finally, for those variables in the long-period variable catalogue \citep{Lebzelter2022} the semi-amplitude, \texttt{amplitude}, has been estimated using a Fourier fit to the epoch photometry. We denote this $\Delta G_\mathrm{Fourier}$. This quantity is only reported if a period has been assigned to the source.

In Fig.~\ref{fig:amp_measure} we show a comparison of the different amplitude measures for the high-amplitude long-period variable sample with $\Delta G_\mathrm{Fourier}$ and $80<P/\,\mathrm{day}<1000$. In general, there is a very good agreement between the different measures. We see that for the stars with periods within $20$ days of $190$ or $380$ day (troublesome periods for Gaia) $\Delta G_\mathrm{Fourier}$ is larger than the other measures. This is due to the clustering of measurements around a small range of phases so any amplitude measurement based on the data is underestimated and any model fit is unconstrained over a wide range of phases and thus can be overestimated. The amplitude measure based on the photometric uncertainties is biased low relative to the Fourier amplitude when there are fewer measurements used in the photometric pipeline \citep{Evans2018} than used in the variable star processing. This is possibly due to the variability of these stars leading to observations being sigma-clipped from the photometric pipeline\footnote{\url{https://gea.esac.esa.int/archive/documentation/GDR3}}. There is also the suggestion that more outliers are removed for sources that fluctuate around the windowing configuration changes. However, in the main, the agreement between the different amplitude measures is very good. Removing sources with periods within $20\,\mathrm{day}$ of $190$ or $380$ day we find the median ratios $\Delta G_\mathrm{std}/\Delta G_\mathrm{Fourier}=1.032$, $\Delta G_\mathrm{range}/\Delta G_\mathrm{Fourier}=1.069$ and  $\Delta G_\mathrm{phot}/\Delta G_\mathrm{Fourier}=1.040$ where we have removed sources with the number of mean photometric measurements less than $95\percent$ the number of measurements used in the variability pipeline.

\bsp	
\label{lastpage}
\end{document}